\documentclass[aps,prc,reprint,twocolumn,amsmath,fleqn,dvipsnames,nofootinbib,superscriptaddress]{revtex4-1}

\usepackage[utf8]{inputenc}
\usepackage[T1]{fontenc}

\usepackage{graphicx}
\usepackage{rotating}
\usepackage{pifont}
\usepackage{xcolor}

\usepackage{scalefnt}

\usepackage{amsmath}
\usepackage{amssymb}
\usepackage{amsfonts}
\usepackage{bm}
\usepackage{xfrac}
\usepackage{braket}

\usepackage{lmodern}
\usepackage{microtype}

\usepackage{hyperref}

\hypersetup{
     colorlinks = true,
     citecolor  = blue,
     linkcolor  = blue,
     urlcolor  = black
}

\usepackage{scalefnt}

\definecolor{indianred}{rgb}{0.86, 0.08, 0.24}
\definecolor{royalblue}{rgb}{0.25, 0.41, 0.88}
\definecolor{darkorange}{rgb}{1.0, 0.55, 0}
\definecolor{mediumseagreen}{rgb}{0.24, 0.70, 0.44}
\definecolor{purple}{rgb}{0.5, 0, 0.5}
\definecolor{cyan3}{rgb}{0, 0.80, 0.80}

\definecolor{plot1}{rgb}{0.86, 0.08, 0.24}
\definecolor{plot2}{rgb}{0.25, 0.41, 0.88}
\definecolor{plot3}{rgb}{1.0, 0.55, 0}
\definecolor{plot4}{RGB}{61,153,86}

\begin{document}

%authors
\title{Zero-pairing limit of Hartree-Fock-Bogoliubov reference states}

\author{T. Duguet}
\email{thomas.duguet@cea.fr}
\affiliation{IRFU, CEA, Universit\'e Paris-Saclay, 91191 Gif-sur-Yvette, France}
\affiliation{KU Leuven, Instituut voor Kern- en Stralingsfysica, 3001 Leuven, Belgium}

\author{B. Bally}
\email{benjamin.bally@uam.es}
\affiliation{Departamento de F\'isica Te\'orica, Universidad Aut\'onoma de Madrid, E-28049 Madrid, Spain}

\author{A. Tichai}
\email{alexander.tichai@physik.tu-darmstadt.de}
\affiliation{Max-Planck-Institut f\"ur Kernphysik, Heidelberg, Germany}
\affiliation{Institut f\"ur Kernphysik, Technische Universit\"at Darmstadt, Darmstadt, Germany}
\affiliation{ExtreMe Matter Institute EMMI, GSI Helmholtzzentrum f\"ur Schwerionenforschung GmbH, Darmstadt, Germany}
\affiliation{ESNT, CEA-Saclay, DRF, IRFU, D\'epartement de Physique Nucl\'eaire, Universit\'e de Paris Saclay, F-91191 Gif-sur-Yvette, France}

\begin{abstract}
\begin{description}
\item[Background] The variational Hartree-Fock-Bogoliubov (HFB) mean-field theory is the starting point of various (\emph{ab initio}) many-body methods dedicated to superfluid systems. In this context, pairing correlations may be driven towards zero either on purpose via HFB calculations  {\it constrained} on, e.g., the particle-number variance or simply because inter-nucleon interactions cannot sustain pairing correlations in the first place in, e.g., closed-shell systems. While taking this limit constitutes a text-book problem when the system is of closed-(sub)shell character, it is typically, although wrongly, thought to be ill-defined whenever the naive filling of single-particle levels corresponds to an open-shell system. 
\item[Purpose] The present work demonstrates that the zero-pairing limit of an HFB state is well-defined independently of the average particle number A it is constrained to. Still, the nature of the limit state is shown to depend of the regime, i.e., on whether the nucleus characterizes as a closed-(sub)shell or an open-shell system when taking the limit. Finally, the consequences of the zero-pairing limit on Bogoliubov many-body perturbation theory (BMBPT) calculations performed on top of the HFB reference state are illustrated. 
\item[Methods] The zero-pairing limit of a HFB state constrained to carry an arbitrary (integer) number of particles A on average is worked out analytically and realized numerically using a two-nucleon interaction derived within the frame of chiral effective field theory.
\item[Results] The zero-pairing limit of the HFB state is mathematically well-defined, independently of the closed- or open-shell character of the system in the limit. Still, the {\it nature} of the limit state strongly depends on the underlying shell structure and on the associated naive filling reached in the zero-pairing limit for the particle number A of interest. First, the text-book situation is recovered for closed-(sub)shell systems, i.e., the limit state is reached for a {\it finite} value of the inter-nucleon interaction (the well-known BCS collapse) and takes the form of a {\it single} Slater determinant displaying (i) {\it zero} pairing energy, (ii) {\it non-degenerate} elementary excitations and (iii) {\it zero} particle-number variance. Contrarily, a non-standard situation is obtained for open-shell systems, i.e., the limit state is only reached for a {\it zero} value of the pairing interaction (no BCS collapse) and takes the form of a specific {\it finite linear combination} of Slater determinants displaying (a) a {\it non-zero} pairing energy, (b) {\it degenerate} elementary excitations and (c) a {\it non-zero} particle-number variance for which an analytical formula is derived. This non-zero particle-number variance acts as a lower bound that depends in a specific way on the number of valence nucleons and on the degeneracy of the valence shell. The fact that a given nucleus does belong to one category or the other, i.e., closed-(sub)shell or open shell, in the pairing limit may depend on the self-consistent spatial symmetry assumed in the calculation. All these findings are confirmed and illustrated numerically. Last but not least, BMBPT calculations of closed-shell (open-shell) nuclei are shown to be well-defined (ill-defined) in the zero-pairing limit.
\item[Conclusions] While HFB theory has been intensively scrutinized formally and numerically over the last decades, it still uncovers unknown and somewhat unexpected features. In the present paper, the zero-pairing limit of a HFB state carrying an arbitrary number of particles has been worked out and shown to lead to drastic differences and consequences depending on the closed-(sub)shell or open-shell nature of the system in that limit. From a general perspective, the present analysis demonstrates that HFB theory does {\it not} reduce to HF theory even when the pairing field is driven to zero in the HFB Hamiltonian matrix. 
\end{description}
\end{abstract}

%\begin{keyword}
%tensor decomposition, many-body theory, ab initio nuclear theory
%\end{keyword}

\maketitle

\section{Introduction}

Based on the use of a single {\it product} state, Hartree-Fock-Bogoliubov (HFB) theory~\cite{RiSc80} provides a variational mean-field approximation method to many-fermion systems capable of tackling pairing correlations responsible for (nuclear) superfluidity. It does so at the price of breaking $U(1)$ global gauge symmetry associated with particle-number conservation. As a result, the HFB solution can only be constrained to carry the correct particle number A on {\it average} and displays a non-zero particle-number dispersion reflecting, i.e., varying with, the amount of pairing correlations in the system. 

While the HFB state captures the essence of static (strong) correlations associated with superfluidity in open-shell nuclei, additional correlations must be incorporated in order to reach a fully quantitative description of the exact, e.g., ground state of $H$, i.e.,
\begin{enumerate}
\item dynamical (weak) correlations can be efficiently captured by expanding the exact ground-state around the HFB reference state in a perturbative fashion, i.e., via Bogoliubov many-body perturbation theory (BMBPT)~\cite{Tichai:2018mll,Arthuis:2018yoo,Tichai:2020dna} or in a non-perturbative way via, e.g., Bogoliubov coupled cluster (BCC)~\cite{Si15} or Gorkov self-consistent Green's function (GSCGF)~\cite{So11,Soma:2013xha} theories. 
\item Given that the HFB state is not an eigenstate of the particle-number operator $A$, it contains, together with the (truncated) expansions built on it, a symmetry contamination. Removing the contaminants by restoring $U(1)$ global gauge symmetry leads to capturing additional static correlations. This can be achieved via projected BMBPT (PBMBPT)~\cite{Duguet:2015yle} or projected BCC (PBCC) theory~\cite{Duguet:2015yle,Qiu:2018edx}. At lowest order, PBMBPT and PBCC reduce to projected HFB (PHFB) theory~\cite{RiSc80}.
\end{enumerate}
In conclusion, the HFB state typically acts as a reference state for more advanced, potentially exact, many-body expansion methods.

It is often casually stated that the HFB state, or any Bogoliubov state for that matter, reduces to a Slater determinant (i) when taking a zero-pairing limit or (ii) when targeting a closed-shell system, typically implying that both statements are essentially equivalent knowing that (iii) the zero-pairing limit is ill-defined for an open-shell system, i.e., it can only be safely considered for a closed-(sub)shell nucleus. Because these statements are only partially correct, the goal of the present contribution is to investigate analytically and illustrate numerically 
\begin{enumerate}
\item the zero-pairing limit of a HFB state constrained to carry an arbitrary particle-number A on average,
\item the behavior of BMBPT in such a limit.
\end{enumerate}

In the present work, the HFB state is shown to reach a mathematically well-defined zero-pairing limit, even for open-shell nuclei. However, the nature and characteristics of that limit state depend strongly on the closed- or open-shell character of the system, i.e., on the nature of the underlying shell structure and of the associated naive filling reached in the zero-pairing limit. Furthermore, these features may themselves depend on the self-consistent spatial symmetry assumed in the calculation. To the best of our knowledge, these basic properties of HFB theory and of the underlying Bogoliubov algebra have never been fully uncovered. Last but not least, the impact of taking the zero-pairing limit on PHFB and BMBPT is further discussed.

The present paper is organized as follows. While Sec.~\ref{basics} introduces the necessary ingredients for the remainder of the paper, Sec.~\ref{zeropairingSec} proceeds to the analytical investigation of the zero-pairing limit. Next, Sec.~\ref{Numresults} displays the results of the numerical calculations illustrating the analytical conclusions reached in the previous section. Eventually, Sec.~\ref{conclusions} provides the conclusions of the present work. A short appendix complements the paper.

\section{Basic ingredients}
\label{basics}

The present section briefly introduces constrained and unconstrained HFB, PHFB and BMBPT formalisms in order to be in position to discuss their zero-pairing limit in Sec.~\ref{zeropairingSec}.

\subsection{Hartree-Fock-Bogoliubov formalism}

\subsubsection{Unconstrained calculations}

The Bogoliubov state  $| \Phi \rangle$ is a vacuum for the set of quasi-particle operators obtained via a unitary linear transformation of the form~\cite{RiSc80}
\begin{subequations}
\begin{align}
\beta_{\nu} &\equiv \sum_p U^*_{p\nu} c_p + V_{p\nu} c^\dagger_p\, , \\
\beta_{\nu}^\dagger &\equiv \sum_p U^*_{p\nu} c^\dagger_p + V_{p\nu} c_p \, .
\end{align}
\end{subequations}
where $\{c^{\dagger}_{p}\}$ ($\{c_{p}\}$) defines the set of creation (annihilation) operators associated with the working basis of the one-body Hilbert space ${\cal H}_1$.

While  $| \Phi \rangle$ is not an eigenstate of the particle-number operator $A$, its expectation value must be constrained to match the number of particles A of the targeted system. This is enforced by adding a Lagrange term to the Hamiltonian, thus introducing the so-called grand potential $\Omega = H - \lambda (A-\text{A})$. The Lagrange multiplier $\lambda$ plays the role of the chemical potential and is to be adjusted so that the particle number is indeed correct on average\footnote{In actual applications, one Lagrange multiplier relates to constraining the neutron number N and one Lagrange multiplier is used to constrain the proton number Z. In our discussion A stands for either one of them.}. In this context, the HFB formalism corresponds to minimizing the expectation value of $\Omega$, i.e., the {\it Routhian},
\begin{align}
\Omega_{| \Phi \rangle}  & \equiv \langle \Phi | \Omega | \Phi \rangle \label{HFBrouthian} \\
&=   \sum_{ij} t_{ij} \, \rho_{ij} + \frac{1}{2} \sum_{ijkl} \overline{v}_{ijkl} \, \rho_{ki} \, \rho_{lj}   \nonumber \\
& \hspace{0.2cm} + \frac{1}{4} \sum_{ijkl} \overline{v}_{ijkl} \, \kappa^{\ast}_{ij} \, \kappa_{kl} -\lambda \big(\sum_{ij} \delta_{ij} \, \rho_{ij}  - \, \text{A}\big) \nonumber \, , 
\end{align}
within the manifold of Bogoliubov states. This procedure delivers the HFB eigenvalue equation~\cite{RiSc80}
\begin{align}
  \label{eq:hfb_equationunconstrained}
\begin{pmatrix} h - \lambda &  \Delta \\ - \Delta^\ast & -(h -\lambda)^\ast \end{pmatrix} \begin{pmatrix} U \\ V \end{pmatrix}_{\mu} 
  &= E_{\mu} \begin{pmatrix} U \\ V \end{pmatrix}_{\mu} \, ,
\end{align}
providing the set of  quasi-particle eigenstates making up the columns of the transformation matrices $(U,V)$ as well as the set of quasi-particle energies $\{E_{\mu}\}$ as eigenvalues. 

In Eq.~\eqref{eq:hfb_equationunconstrained}, the Hartree-Fock and Bogoliubov fields depend on matrix elements of the one-body kinetic energy operator $\{t_{ij}\}$ and of the two-body interaction\footnote{In the present investigation, original and induced three-nucleon forces are omitted for simplicity given that none of the conclusions depend on their inclusion. When  taking them into consideration, the particle-number conserving normal-ordered two-body (PNO2B) approximation introduced in Ref.~\cite{Ripoche:2019nmy} can be used to take the dominant part of three-body forces into account via an effective two-body-like interaction as was done in, e.g., Ref.~\cite{Tichai:2018mll}.} operator  $\{\overline{v}_{ikjl}\}$ according to\footnote{One subtracts the center-of-mass kinetic energy to the Hamiltonian $H$ in actual calculations of finite nuclei. As far as the present work is concerned, this simply leads to a redefinition of the one- and two-body matrix elements $t_{ij} $ and $\overline{v}_{ikjl}$ of the Hamiltonian without changing any aspect of the analysis.}
\begin{subequations}
\label{fields}
\begin{align}
h_{ij} &\equiv t_{ij} + \sum_{kl} \overline{v}_{ikjl} \, \rho_{lk}   \, , \label{field1} \\
\Delta_{ij} &\equiv \frac{1}{2} \sum_{kl} \overline{v}_{ijkl} \, \kappa_{kl}  
\, , \label{field2} 
\end{align}
\end{subequations}
where
\begin{subequations}
\label{densities}
\begin{align}
\rho_{ij} &\equiv  \langle \Phi | c_{j}^{\dagger}  c_{i} | \Phi \rangle  = \sum_{\nu} V^{\ast}_{i\nu} V_{j\nu} , \\
\kappa_{ij} &\equiv \langle \Phi | c_{j}  c_{i} | \Phi \rangle = \sum_{\nu}  V^{\ast}_{i\nu} U_{j\nu} \, ,
\end{align}
\end{subequations}
respectively denote the normal and anomalous density matrices associated with  $| \Phi \rangle$. When building the density matrices in Eq.~\eqref{densities} from the solutions of Eq.~\eqref{eq:hfb_equationunconstrained} , the sum is actually restricted to quasi-particle states associated with positive quasi-particle energies $\{E_{\mu} \geq 0\}$; i.e., the fully paired vacuum carrying even-number parity is considered throughout the present paper.

Once Eq.~\eqref{eq:hfb_equationunconstrained} is solved, the HFB state can be most conveniently written in its canonical, i.e., BCS-like, form~\cite{RiSc80}
\begin{equation}
| \Phi \rangle \equiv \prod_{k>0} \left[u_k + v_k a^{\dagger}_k a^{\dagger}_{\bar{k}}\right] | 0 \rangle \, . \label{HFBstate}
\end{equation}
In Eq.~\eqref{HFBstate}, operators $\{ a^{\dagger}_k, a_k\}$ characterize the so-called canonical one-body basis in which pairs of conjugated states $(k,\bar{k})$ are singled out by the Bogoliubov transformation. Conventionally, the two members of the conjugated pair are distinguished as $k>0$ and $\bar{k}<0$, thus, effectively splitting the basis into two halves. The coefficients $u_k=+u_{\bar{k}}$ and $v_k=-v_{\bar{k}}$ are BCS-like occupation numbers. They make up the simplified Bogoliubov transformation obtained through the Bloch-Messiah-Zumino decomposition~\cite{RiSc80} of the full Bogoliubov transformation extracted from Eq.~\eqref{eq:hfb_equationunconstrained}. The BCS-like occupation numbers can be chosen real and satisfy the identity $u^2_k + v^2_k = 1$. 

Employing Eq.~\eqref{HFBstate}, the canonical form of the norm and density matrices of the HFB state are obtained as
\begin{equation}
\langle \Phi | \Phi \rangle = \prod_{k>0} (u^2_k + v^2_k) = 1 \, , \label{norm}
\end{equation}
and
\begin{subequations}
\label{densitiescano}
\begin{align}
\rho_{kk'} &= v^2_k  \, \delta_{kk'}, \\
\kappa_{kk'} &= u_kv_k \, \delta_{\bar{k}k'},
\end{align}
\end{subequations}
respectively.

Thanks to an appropriate adjustment of the chemical potential $\lambda$ in Eq.~\eqref{eq:hfb_equationunconstrained}, the average particle number carried by the HFB state is constrained to the integer value A, which in the canonical basis reads as
\begin{align}
\langle \Phi | A | \Phi \rangle  & = \sum_{k} v_k^2 \nonumber \\
&\equiv \sum_{k} \frac{1}{2} \left(1 - \frac{\epsilon_k - \lambda}{\sqrt{(\epsilon_k - \lambda)^2 + \Delta_k^2}}\right) \nonumber \\
&= \text{A} \, , \label{partnumbconstr}
\end{align}
where $\epsilon_k \equiv h_{kk} = h_{\bar{k}\bar{k}}$ and $\Delta_k \equiv \Delta_{k\bar{k}} = - \Delta_{\bar{k}k}$.

Eventually, the total HFB energy  is obtained as
\begin{align}
E_{| \Phi \rangle}  & \equiv \langle \Phi | H | \Phi \rangle \nonumber \\
& \equiv E^{\text{kin}}_{| \Phi \rangle} + E^{\text{HF}}_{| \Phi \rangle} + E^{\text{B}}_{| \Phi \rangle} \nonumber \\
&= \sum_{ij} t_{ij} \, \rho_{ij} + \frac{1}{2} \sum_{ijkl} \overline{v}_{ijkl} \, \rho_{ki} \, \rho_{lj}  \nonumber \\
& \hspace{2cm}  + \frac{1}{4} \sum_{ijkl} \overline{v}_{ijkl} \, \kappa^{\ast}_{ij} \, \kappa_{kl} \nonumber \\
&= \sum_{k} t_{kk} \, v_k^2  + \frac{1}{2} \sum_{kk'} \overline{v}_{kk'kk'} \, v_k^2  \, v_{k'}^2  \nonumber \\
& \hspace{2cm} +  \frac{1}{4} \sum_{kk'} \overline{v}_{k\bar{k}k'\bar{k}'} \, u_{k}v_{k} \, u_{k'}v_{k'}  \, , \label{HFBenergy}
\end{align}
and is equal to the Routhian $\Omega_{| \Phi \rangle}$ (Eq.~\eqref{HFBrouthian}) as long as the constraint on the average particle number is indeed satisfied.

\subsubsection{Constrained calculations}

The zero-pairing limit is to be achieved by constraining the {\it variational determination} of the HFB state, i.e., by subtracting from the grand potential $\Omega$ a Lagrange term proportional to an appropriate operator $O$ in such a way that {\it the pairing field is entirely driven to zero in the resulting HFB Hamiltonian matrix}. Once the constrained HFB state is obtained, associated many-body observables, e.g., the binding energy and particle-number variance, can be computed.

Any quantity $O$ varying monotonically with the amount of pairing correlations carried by the HFB state, i.e., acting as an order parameter of the breaking of $U(1)$ global-gauge symmetry, can be employed as a constraint. A typical example is the two-body operator associated with the particle-number variance $(A- \langle \Phi | A | \Phi \rangle)^2$~\cite{siegal72a,faessler73a,faessler75a,meyer91a,Fernandez:2005ux,Bender:2006tb,Vaquero:2011hq,Vaquero:2013paa}. In the present paper, the Hermitian one-body particle-number non-conserving operator
\begin{align}
  \label{constraint}
\Delta_{\text{C}} &\equiv \frac{1}{2} \sum_{ij} \Delta_{ij} \, c^{\dagger}_i c^{\dagger}_j + \frac{1}{2} \sum_{ij} \Delta^{\ast}_{ij} \, c_j c_i  \, ,
\end{align}
with $\Delta_{ij}$ defined in Eq.~\eqref{field2}, is employed. With this operator at hand, $\Omega$ is replaced by the {\it constrained} grand potential
\begin{equation}
\Omega(\delta)\equiv \Omega - \frac{1}{2}(1-\delta) \Delta_{\text{C}} 
\end{equation}
to perform the minimization within the manifold of even-number parity Bogoliubov states. This leads to minimizing the modified, i.e., constrained, Routhian
\begin{align}
\Omega(\delta)_{| \Phi \rangle}  & \equiv \langle \Phi | \Omega(\delta) | \Phi \rangle \label{HFBrouthianconstrained}  \\
&=   \sum_{ij} t_{ij} \, \rho_{ij} + \frac{1}{2} \sum_{ijkl} \overline{v}_{ijkl} \, \rho_{ki} \, \rho_{lj}   \nonumber \\
& \hspace{0.2cm} + \frac{\delta}{4} \sum_{ijkl} \overline{v}_{ijkl} \, \kappa^{\ast}_{ij} \, \kappa_{kl} -\lambda \big(\sum_{ij} \delta_{ij} \, \rho_{ij}  - \, \text{A}\big) \nonumber \, , 
\end{align}
which takes the same form as the unconstrained Routhian, except that the pairing, i.e., Bogoliubov, term is now rescaled by the parameter $\delta$. Of course, constrained and unconstrained Routhians match for $\delta = 1$.

The minimization of $\Omega(\delta)_{| \Phi \rangle}$ leads to solving a constrained HFB eigenvalue equation taking the form\footnote{While providing the same end results as the particle-number variance, the constraining one-body operator $\Delta_{\text{C}}$ is gentler numerically and allows one to reach the zero-pairing limit in a controlled fashion. The numerical easiness is also largely due to the constraining method employed here. Instead of using an actual Lagrange method in which the driving parameter $\delta$ is self-consistently adjusted to make the constraint $\langle \Phi(\delta)| \Delta_{\text{C}}| \Phi(\delta) \rangle$ equate a set of predefined values, the calculation is performed for a fixed value of $\delta$ and is repeated such that $\delta$ scans a chosen interval $[0,\delta_{\text{max}}]$. This approach is appropriate because (i) the specific value of the quantity $\langle \Phi(\delta)| \Delta_{\text{C}}| \Phi(\delta) \rangle$ is of no particular interest and because (ii) the particle-number variance varies monotonically with $\delta$ such that the end results can anyway be displayed as a function of it. In this context, $\delta_{\text{max}}$ can always be taken large enough to cover any desired range of particle-number variance values.}
\begin{align}
  \label{eq:hfb_equation}
\begin{pmatrix} h - \lambda & \delta \times \Delta \\ -\delta \times \Delta^\ast & -(h -\lambda)^\ast \end{pmatrix}_{(\delta)} \begin{pmatrix} U(\delta) \\ V(\delta) \end{pmatrix}_{\mu} 
  &= E_{\mu}(\delta) \begin{pmatrix} U(\delta) \\ V(\delta) \end{pmatrix}_{\mu} \, ,
\end{align}
under the additional constraint that the solution, denoted as $| \Phi(\delta) \rangle$, carries the average nucleon number $\text{A}$. This procedure delivers $\delta$-dependent quasi-particle states and energies $\{E_{\mu}(\delta)\}$, and thus a $\delta$-dependent many-body state $| \Phi(\delta) \rangle$. In Eq.~\eqref{eq:hfb_equation}, the Hartree-Fock and Bogoliubov fields themselves depend on $\delta$ through their functional dependence on the normal and anomalous density matrices associated with $| \Phi(\delta) \rangle$. As visible in Eq.~\eqref{eq:hfb_equation}, the use of $\Omega(\delta)$ eventually boils down to the fact that the pairing field at play in the HFB matrix is obtained by multiplying the unconstrained one by the parameter $\delta$, which is itself effectively equivalent to rescaling all two-body matrix elements entering the pairing field (Eq.~\eqref{field2}) by that same factor. Whereas $\delta =1$ corresponds to the unconstrained calculation, taking $\delta \rightarrow 0$ characterizes the zero-pairing limit of present interest.

Eventually, all quantities introduced in the context of unconstrained calculations can be similarly defined here at the price of providing them with a $\delta$ dependence. The expression of the constrained HFB energy $E_{| \Phi (\delta) \rangle}$ is formally identical to the unconstrained one given in Eq.~\eqref{HFBenergy} except that it acquires an implicit dependence on $\delta$ through the density matrices of the constrained HFB state $| \Phi (\delta) \rangle$. This is at variance with the constrained Routhian $\Omega(\delta)_{| \Phi (\delta) \rangle}$ and the pairing field matrix elements $\Delta_{ij}(\delta)$ that additionally carry an {\it explicit} dependence on $\delta$ in their very definition. In any case, the $\delta$ dependence of the various quantities at play is omitted for simplicity in the remainder of the paper, except if specified otherwise.

\subsection{Bogoliubov many-body perturbation theory}
\label{BMBPTformalism}

One of the focus of the present study is to investigate the consequence of driving the HFB state to the zero-pairing limit when performing a BMBPT calculations on top of it. This investigation happens to raise non-trivial questions regarding the way a perturbative expansion is best formulated when performed on top of a reference state delivered via a {\it constrained} minimization. 

\subsubsection{Unconstrained HFB reference state}
\label{BMBPTunconstrained}

To be in position to address these questions, let us first briefly recall the main ingredients of BMBPT  based on an {\it unconstrained} HFB reference state. For a detailed account of the BMBPT formalism, the reader is referred to Ref.~\cite{Arthuis:2018yoo}.

Because the HFB reference state is not an eigenstate of $A$, the operator meaningfully driving the BMBPT expansion\footnote{The "driving" operator is at play in the (imaginary) time evolution operator transforming the HFB reference state into the fully correlated ground-state and that is Taylor expanded to build the perturbative series~\cite{Duguet:2015yle}.} is not $H$ but $\Omega$~\cite{Duguet:2015yle} and is thus the same as the one at play in the HFB minimization. To set up BMBPT, $\Omega$ must be first normal ordered with respect to  $| \Phi \rangle$
\begin{align}
\Omega &=   \Omega^{[0]}_{| \Phi \rangle} + \Omega^{[2]}_{| \Phi \rangle} + \Omega^{[4]}_{| \Phi \rangle} \notag \\
&= \Omega^{00}_{| \Phi \rangle}   \notag \\
&\phantom{=} + 
 \Omega^{20}_{| \Phi \rangle} + \Omega^{11}_{| \Phi \rangle} +\Omega^{02}_{| \Phi \rangle} \notag \\
&\phantom{=} +   \Omega^{40}_{| \Phi \rangle} + \Omega^{31}_{| \Phi \rangle} +\Omega^{22}_{| \Phi \rangle} +\Omega^{13}_{| \Phi \rangle} +\Omega^{04}_{| \Phi \rangle} \, ,
\label{eq:NO}
\end{align} 
where $\Omega^{ij}_{| \Phi \rangle}$ denotes the normal-ordered component involving $i$ ($j$) quasi-particle creation (annihilation) operators associated with $| \Phi \rangle$, e.g., 
\begin{align}
\Omega^{31}_{| \Phi \rangle} &\equiv \frac{1}{3!}\sum_{\mu_1 \mu_2 \mu_3 \mu_4}  \Omega^{31}_{\mu_1 \mu_2 \mu_3 \mu_4}
   \beta^{\dagger}_{\mu_1}\beta^{\dagger}_{\mu_2}\beta^{\dagger}_{\mu_3}\beta_{\mu_4} \, .
\end{align} 
In Eq.~\eqref{eq:NO}, $\Omega^{00}_{| \Phi \rangle}$ is nothing but the Routhian $\Omega_{| \Phi \rangle}$ introduced in Eq.~\eqref{HFBrouthian}, $\Omega^{[2]}_{| \Phi \rangle}$ is an effective, i.e., normal-ordered, one-body operator and $\Omega^{[4]}_{| \Phi \rangle}$ is an effective two-body one. Details on the normal-ordering procedure as well as expressions of the matrix elements of each operator $\Omega^{ij}_{| \Phi \rangle}$ in terms of the original matrix elements of the Hamiltonian and of the $(U,V)$ matrices can be found in Ref.~\cite{Si15}.

To actually set up the perturbation theory, the grand potential is split into an unperturbed part $\Omega_{0}$ and a residual part $\Omega_1$
\begin{equation}
\label{split1}
\Omega = \Omega_{0} + \Omega_{1} \ ,
\end{equation}
such that
\begin{subequations}
\label{split2}
\begin{align}
\Omega_{0} &\equiv \Omega^{00}_{| \Phi \rangle}+\tilde{\Omega}^{11}_{| \Phi \rangle; \{\tilde{E}_{\mu}\}} \ , \\
\Omega_{1} &\equiv \Omega^{20}_{| \Phi \rangle} + \breve{\Omega}^{11}_{| \Phi \rangle; \{\tilde{E}_{\mu}\}} + \Omega^{02}_{| \Phi \rangle} \notag \\
&\phantom{=} +   \Omega^{40}_{| \Phi \rangle} + \Omega^{31}_{| \Phi \rangle} +\Omega^{22}_{| \Phi \rangle} +\Omega^{13}_{| \Phi \rangle} +\Omega^{04}_{| \Phi \rangle} \ ,
 \label{e:perturbation}
\end{align}
\end{subequations}
with $\breve{\Omega}^{11}_{| \Phi \rangle; \{\tilde{E}_{\mu}\}}\equiv\Omega^{11}_{| \Phi \rangle}- \tilde{\Omega}^{11}_{| \Phi \rangle; \{\bar{E}_k\}}$. The one-body part of $\Omega_{0}$ is diagonal, i.e.,
\begin{equation}
 \tilde{\Omega}^{11}_{| \Phi \rangle; \{ \tilde{E}_{\mu}\}} \equiv \sum_{\mu} \tilde{E}_{\mu} \, \beta^{\dagger}_{\mu} \beta_{\mu} \, , \label{onebodypiece}
\end{equation}
with $\{\tilde{E}_{\mu}\}$ denoting an arbitrary set of positive energies. 

As $|\Phi \rangle$ solves the unconstrained HFB variational problem, i.e., Eq.~\eqref{eq:hfb_equationunconstrained}, one has that $\Omega^{20}_{| \Phi \rangle}=\Omega^{02}_{| \Phi \rangle}=0$. Furthermore, while the choice of $\tilde{\Omega}^{11}_{| \Phi \rangle; \{\tilde{E}_{\mu}\}}$, i.e., of the set of energies $\{\tilde{E}_{\mu}\}$, is arbitrary, a natural choice is to pick the eigenvalues of the HFB eigenvalue equation, i.e., to choose $\tilde{E}_{\mu}\equiv E_{\mu} > 0$ for all $\mu$ in Eq.~\eqref{onebodypiece}. This choice additionally leads to $\breve{\Omega}^{11}_{| \Phi \rangle; \{E_{\mu}\}}=0$ such that the residual interaction $\Omega_1$ in Eq.~\eqref{e:perturbation} reduces to its effective two-body part $\Omega^{[4]}_{| \Phi \rangle}$. This particular setting defines the \emph{canonical} version of BMBPT and reduces significantly the number of non-zero diagrams to be considered. Contrarily, not making such a choice leads to the appearance of \emph{non-canonical} diagrams involving $\Omega^{20}_{|\Phi \rangle}$, $\breve{\Omega}^{11}_{| \Phi \rangle; \{\tilde{E}_{\mu}\}}$ and $\Omega^{02}_{|\Phi \rangle}$ vertices.

The power of BMBPT relies on the fact that the superfluid character of open-shell nuclei ensures that the HFB reference state is non-degenerate, i.e., elementary quasi-particle excitations of the HFB vacuum display non-zero energies. This key property relates to the fact that the quasi-particle energies $\{E_{\mu}\}$ are bound from below by the superfluid \emph{pairing gap} at the Fermi energy
\begin{align}
\text{Min}_{\mu} \{E_{\mu}\} \geq \Delta_{\text{F}} > 0 \, , \label{fermigap}
\end{align}
when the system is indeed superfluid, i.e., exhibits pairing correlations. The benefit if this feature can be best appreciated by considering as an example the first BMBPT correction that, added to the reference HFB energy~\cite{Arthuis:2018yoo}, defines second-order BMBPT calculations, i.e., BMBPT(2),  
\begin{align}
E^{(2)}_{| \Phi \rangle; \{\tilde{E}_{\mu}\}} =& - \frac{1}{2} \sum_{\mu_1 \mu_2} \frac{H^{20}_{\mu_1 \mu_2} \Omega^{02}_{\mu_1 \mu_2} }{\tilde{E}_{\mu_1}+\tilde{E}_{\mu_2}} \label{BMBPTcorrection2} \\
&-\frac{1}{4!} \sum_{\mu_1 \mu_2 \mu_3 \mu_4} \frac{H^{40}_{\mu_1\mu_2\mu_3\mu_4} \Omega^{04}_{\mu_3 \mu_4 \mu_1\mu_2} }{\tilde{E}_{\mu_1}+\tilde{E}_{\mu_2}+\tilde{E}_{\mu_3}+\tilde{E}_{\mu_4}}  \, .   \nonumber
\end{align}
While the first term in Eq.~\eqref{BMBPTcorrection2} is non-canonical and thus cancels in the present context, using $\tilde{E}_{\mu}\equiv E_{\mu} > 0$ leads to strictly positive energy denominators. As $E^{(2)}$ is representative of all correction terms, the expansion is in this case, if not necessarily convergent~\cite{Demol:2020mzd,Tichai:2020dna},  at least ensured to be non singular in open-shell nuclei. This would not be the case in standard MBPT due to the degenerate character of the Hartree-Fock (HF) Slater determinant reference state in open-shell nuclei, i.e., elementary particle-hole excitations of the Slater determinant within the valence shell are zero in such a situation. 

\subsubsection{Constrained HFB reference state}
\label{BMBPTconstrained}

Because the reference state $| \Phi(\delta) \rangle$ is now obtained by solving the constrained HFB eigenvalue equation, the operator $\Omega$ driving the BMBPT expansion differs from the one, i.e., $\Omega(\delta)$, at play in the HFB minimisation. This results in the fact that 
\begin{itemize}
\item the normal-ordered form of $\Omega$ with respect to $| \Phi(\delta) \rangle$ is not canonical\footnote{The operator $\Omega(\delta)$ is in canonical form when normal ordered with respect to  $| \Phi(\delta) \rangle$ but  $\Omega$ is not, except for $\delta =1$ of course.}, i.e., $\Omega^{20}_{| \Phi(\delta) \rangle}$ and $\Omega^{02}_{| \Phi(\delta) \rangle}$ are not zero,
\item the partitioning of $\Omega$, i.e., Eqs.~\eqref{split1}-\eqref{split2} associated with the choice of $\{\tilde{E}_{\mu}(\delta)\}$ in the definition of $\tilde{\Omega}^{11}_{| \Phi(\delta) \rangle; \{\tilde{E}_{\mu}(\delta)\}}$ is neither natural nor obvious. 
\end{itemize}
Thus, the application of BMBPT on top of a constrained HFB state necessarily requires to evaluate non-canonical diagrams. As for the partitioning of $\Omega$, two choices of quasi-particle energies are presently tested at any given value of $\delta$
\begin{enumerate}
\item $\tilde{E}_{\mu}(\delta) = E_{\mu}(\delta)$,
\item $\tilde{E}_{\mu}(\delta)  = E_{\mu} = E_{\mu}(1)$ ,
\end{enumerate}
where the second choice is thus independent of $\delta$. These two options are respectively denoted as {\it Option 1} (BMBPT-1) and  {\it Option 2} (BMBPT-2) in the remainder of the paper. Of course, both options coincide for unconstrained calculations, i.e., for $\delta=1$.

\subsection{Projected Hartree-Fock-Bogoliubov formalism}

The PHFB formalism invokes gauge-rotated HFB states obtained as
\begin{align}
| \Phi(\varphi) \rangle &\equiv R(\varphi)| \Phi \rangle  \notag \\
&= \prod_{k>0} (u_k + e^{2i\varphi}v_k a^{\dagger}_k a^{\dagger}_{\bar{k}}) | 0 \rangle \, , 
\end{align}
where the rotation operator spanning the $U(1)$ group is given by $R(\varphi)\equiv e^{iA\varphi}$, with $\varphi \in [0,2\pi]$. The off-diagonal norm kernel between the HFB state and any gauge-rotated partner generalizes the norm overlap of Eq.~\eqref{norm} according to
\begin{equation}
\langle \Phi | \Phi(\varphi) \rangle = \prod_{k>0} (u^2_k + e^{2i\varphi}v^2_k) \, .  \label{normoverlap}
\end{equation}
One notices that $\langle \Phi| \Phi(\pi/2)  \rangle = 0$ whenever a specific shell is such that $u^2_{k}=v^2_{k}=1/2$. Whenever the corresponding shell is characterized by $p_k$ conjugated pairs, the gauge-dependent integrand at play in the PHFB calculation of an observable associated with a $(p_k\!+\!1)$-body (or higher-body) operator displays an apparent pole~\cite{tajima92a,donau98,almehed01a,anguiano01b}. While this pole is in fact a mere intermediate artefact and disappears when combining the various terms contributing to the observable, it can generate numerical difficulties\footnote{If the operator at play is only of $p_k$-body character, there is no apparent pole but a zero-over-zero when multiplying the off-diagonal norm kernel with the connected off-diagonal operator kernel that can still lead to numerical difficulties. When the operator is of even lower rank, no difficulty arises.} in applications if not accurately resolved~\cite{almehed01a,anguiano01b,doba05a,Bender:2008rn}.

\section{Zero-pairing limit}
\label{zeropairingSec}

With the ingredients of Sec.~\ref{basics} at hand, the goal is now to actually investigate the even-number parity solution of HFB equations zero-pairing limit. 

\subsection{Naive filling}

The discussion below crucially relies on the {\it naive filling} characterizing a given system of interest {\it in the zero-pairing limit}. The naive filling corresponds to occupying single-particle canonical states characterized by the A lowest energies $\epsilon_k$. Doing so, one exhausts the A nucleons in such a way that $0\leq a_v \leq d_v$ nucleons sit in the so-called {\it valence}, i.e., last occupied, shell characterized by energy $\epsilon_v$ and degeneracy $d_v$ (and thus $p_v\equiv d_v/2$ pairs of conjugated states). The naive occupation of each canonical state belonging to the valence shell denoted as
\begin{equation}
o_v \equiv  \frac{a_v}{d_v} \, ,
\end{equation}
ranges between 0 and 1, i.e., $0<o_v \leq 1$. 

Two crucially different categories of nuclei emerge in this context, i.e., a nucleus is either of closed-(sub)shell character when $o_v=1$ or of open-shell character whenever $0<o_v < 1$. In the present context, the definition of these two categories must be understood in a broad sense, i.e., independently of the symmetries, and thus of the degeneracies, characterizing the spectrum $\{\epsilon_k\}$. While the fact that a given nucleus belongs to one category or the other can only be inferred \textit{a posteriori} and will depend on the symmetries characterizing the employed numerical code\footnote{A system whose HFB solution is of open-shell character in the zero-pairing limit whenever spherical symmetry is enforced can relax to a closed-shell system if $SU(2)$ rotational symmetry is allowed to break.}, the general features of the zero-pairing limit will not depend on the fact that the closed-shell/open-shell system is spherical or deformed. 

Whenever the canonical energy shells $\{\epsilon_k\}$ obtained in the zero-pairing limit is characterized by spherical symmetry, they each display a degeneracy $d_k \equiv 2j_k+1$ where $j_k$ denotes the one-body total angular momentum shared by all degenerate single-particle states. For example, the text-book expectation is that $^{26}$O respects spherical symmetry and is characterized by a $\text{d}_{3/2}$ valence shell carrying degeneracy $d_v=4$ and fitting 2 neutrons such that $o_v =1/2$. Whenever $SU(2)$ rotational symmetry is allowed to break, the canonical spectrum  $\{\epsilon_k\}$ associated with the ground-state of any even-even nucleus qualifying as a {\it spherical} doubly open-shell is only left with the two-fold Kramers degeneracy such that these nuclei all eventually qualify as {\it deformed} closed-shell systems\footnote{As will be exemplified in the numerical applications, such a situation also occurs for semi-magic nuclei.}.

\subsection{Definition of the limit}

Except when the naive filling reached in the zero-pairing limit corresponds to a closed-(sub)shell system, i.e., whenever $o_v=1$, $| \Phi \rangle$ cannot reduce to a Slater determinant. Thus, open-shell nuclei characterized by $0<o_v<1$  constitute the non-trivial focus of the present study. Of course, the higher the degree of symmetry, i.e., the degenerate character of single-particle energy shells, the larger the occurrence rate of open-shell systems. For such nuclei, the zero-pairing limit must be formally defined and performed with care. How the zero-pairing limit is taken is important and non-trivial because the HFB state
\begin{enumerate}
\item must be constrained to fulfilling Eq.~\eqref{partnumbconstr},
\item is solution of the iterative HFB eigenvalue problem (Eq.~\eqref{eq:hfb_equation}) involving an interference between the Hartree-Fock and the Bogoliubov fields\footnote{When explicitly performing the constraint on the particle-number variance, the constrained can eventually be shared at will between both fields entering the HFB matrix via the use of identities deriving from the unitarity of the Bogoliubov transformation~\cite{Ripochethesis2019}. If $100\%$ of the constraint is injected into the Hartree-Fock field, the constraint induces more and more spread out nuclear shells that makes the pair scattering less and less efficient. Contrarily, when $100\%$ of the constraint is injected into the Bogoliubov field, the effective strength of the pairing field is reduced. While observable, e.g., the HFB total energy, and the eigenstates of the HFB matrix are independent of this partitioning, the quasi-particle energies are not. The present way of directly scaling the Bogoliubov field is close in spirit to the second case.}.
\end{enumerate}
As the zero-pairing limit ($\delta \rightarrow 0$) is taken by scaling the Bogoliubov field down to zero in Eq.~\eqref{eq:hfb_equation}, the search for the limit reached by $| \Phi(\delta) \rangle$ expressed in its canonical basis can be analytically materialized by
\begin{equation}
\Delta_k \longrightarrow 0 \,\, \forall k \,\,\, \text{subject to} \,\,\, \langle \Phi | A | \Phi \rangle  = \text{A} \label{limit} \, .
\end{equation}
Applying Eq.~\eqref{limit} in a meaningful fashion leads to distinguishing three categories of canonical single-particle states, i.e., states characterized by
\begin{enumerate}
\item $\epsilon_k - \lambda <0$, casually denoted as "hole states",
\item $\epsilon_k - \lambda =0$, casually denoted as "valence states",
\item $\epsilon_k - \lambda >0$, casually denoted as "particle states",
\end{enumerate}
when reaching the limit. Valence states, which can only concern one shell, must be explicitly considered in order to satisfy Eq.~\eqref{limit} under the assumption that $0<o_v\leq 1$. 

When driving the system towards the limit, the canonical basis changes with $\delta$ such that not only the chemical potential $\lambda$ but also the location of the shells evolve. Under the hypothesis that the spatial symmetry and the associated degeneracies remain unchanged along the way, one cannot exclude (i) the occurrence of shell crossings or (ii) the occurrence/lifting of an accidental degeneracy. Furthermore, if the numerical code allows the system to break spherical symmetry, one also cannot exclude a change of spatial symmetry along the constraining path and, thus, in the limit. As a result, one cannot exclude that the effective degeneracy of the valence shell, and thus the closed- or open-shell character of the associated nucleus, may be different in the zero-pairing limit and in, e.g., the unconstrained calculation. In any case, the hole, valence or particle character of the shells, as well as the associated naive filling, relevant to the present analysis are the ones reached {\it in the zero-pairing limit}.

\subsection{Closed-(sub)shell system}
\label{closedshell}

For reference, let us first study closed-(sub)shell systems. In this case, one can arbitrarily defines the valence shell to be the last fully occupied ($o_v=1$) or the first fully empty ($o_v=0$) shell when proceeding to the naive filling. While the first choice is presently made here, general formulae derived later can be used with both conventions. 

Whenever $SU(2)$ symmetry is self-consistently satisfied, closed-(sub)shell systems are obtained each time a spherical shell is fully occupied when proceeding to the naive filling, e.g., in the semi-magic $^{22}$O for which the neutron 1d$_{5/2}$ shell is fully occupied. Consequently, only a small subset of semi-magic nuclei do (potentially) belong to this category. When relaxing $SU(2)$ rotational symmetry, all even-even nuclei that are not already of spherical doubly-closed-(sub)shell nature tend to deform in the zero-pairing limit to acquire such a character as a result of the residual two-fold Kramers degeneracy and thus display $o_v=1$. 

In this situation, no specific surprise occurs. Equation~\eqref{partnumbconstr} can be trivially fulfilled in the zero-pairing limit by fully occupying (emptying) the A lowest (remaining) canonical single-particle states such that
\begin{equation}
\lim_{\underset{\langle A  \rangle  = \text{A}}{\Delta_k \to 0}} v^{2}_k = 1 (0) \, . \label{limit2}
\end{equation}
Equation~\eqref{limit2} stipulates that no genuine valence shell emerges through the zero-pairing limit and that all single-particle states converge towards either a hole or a particle state. Consequently, the HFB state itself converges trivially to the closed-(sub)shell Slater determinant
\begin{align}
| \bar{\Phi} \rangle & \equiv \lim_{\underset{\langle A  \rangle  = \text{A}}{\Delta_k \to 0 \,\, \forall k}}   | \Phi \rangle  \nonumber \\
&=  \prod_{h=1}^{A/2} a^{\dagger}_h a^{\dagger}_{\bar{h}}| 0 \rangle \, , \label{HFBstatelimit0}
\end{align}
which is an eigenstate of $A$ with eigenvalue A and zero particle-number variance
\begin{align}
\text{VAR}_{| \bar{ \Phi} \rangle} &\equiv \langle \bar{ \Phi} | (A - \langle  \Phi |A | \Phi \rangle)^2| \bar{ \Phi} \rangle \nonumber \\
&= \langle \bar{ \Phi} | A^2 | \bar{\Phi} \rangle - \langle \bar{ \Phi} | A | \bar{ \Phi} \rangle^2   \nonumber \\
&= 0 \, .
\end{align}
Correspondingly, the HFB energy (Eq.~\eqref{HFBenergy}) takes the standard mean-field, i.e., HF, form associated with a Slater determinant 
\begin{align}
\bar{E}  & = \sum_{k=1}^{\text{A}} t_{kk}  + \frac{1}{2} \sum_{kl=1}^{\text{A}} \overline{v}_{klkl}  \, , \label{HFBenergylimit1}
\end{align}
where the Bogoliubov term, i.e., the pairing energy, is strictly zero. 

\subsection{Open-shell system ($o_v=1/2$)}
\label{specific}

Let us continue with the simplest non-trivial case where the naive valence fractional occupation is $o_v=1/2$. The next section will address the general case. Dealing with even particle numbers and spherical symmetry, $o_v=1/2$ corresponds to a half-filled valence shell, e.g., to 2 nucleons sitting in a $\text{p}_{3/2}$ or $\text{d}_{3/2}$ valence shell, 4 nucleons sitting in a $\text{f}_{7/2}$ or $\text{g}_{7/2}$ valence shell, etc. This situation is also encountered when one nucleon sits in a doubly-degenerate valence shell. This occurs for odd-even nuclei whenever enforcing spherical symmetry and whenever a s shell lies at the Fermi energy as well as for any odd-even deformed system whose even-number parity vacuum is characterized by canonical single-particle states displaying two-fold Kramers degeneracy\footnote{These Bogoliubov states correspond to the zero-pairing limit of the fully-paired even number-parity vacua appropriate to odd systems discussed at length in Ref.~\cite{Duguet:2001gr}.}.

To conduct the present discussion, let us consider the simplest situation of two nucleons eventually sitting in a $\text{d}_{3/2}$ valence shell. Based on a text-book spherical single-particle spectrum, this situation is expected to encountered in, e.g., $^{26}$O. It corresponds to having $a_v=2$ and $d_v=4$. The $p_v=2$ pairs of conjugate valence states, generically denoted as $(v,\bar{v})$, are presently specified as $(v_1,v_{\bar{1}})$ and $(v_2,v_{\bar{2}})$. 

The zero-pairing limit of hole and particle states works as in Eq.~\eqref{limit2} with $\text{A}-2$, i.e., 24, particles eventually occupying hole states. For valence states, the situation is more subtle. To reach the average occupation $o_v=1/2$ in the limit in a controlled fashion, one must assume\footnote{This property is validated numerically in Sec.~\ref{Numresults}.} that
\begin{equation}
\lim_{\underset{\langle A  \rangle  = \text{A}}{\Delta_{v} \to 0}} \left|\frac{\epsilon_{v} - \lambda}{\Delta_{v}}\right| = 0 \, , \label{limit3}
\end{equation}
i.e., that $(\epsilon_{v} - \lambda)$ goes to 0 faster than $\Delta_{v}$. With Eq.~\eqref{limit3} at hand, one indeed obtains the required limit for the average single-particle valence state occupation
\begin{equation}
\lim_{\underset{\langle A  \rangle  = \text{A}}{\Delta_{v} \to 0}} v^{2}_{v} = \frac{1}{2} = o_v \, . \label{limit4}
\end{equation}
Together with Eq.~\eqref{limit2} for 24 particles, Eq.~\eqref{limit4} allows one to fulfill Eq.~\eqref{partnumbconstr} for $\text{A}=26$ in the zero-pairing limit. 

With this carefully performed limit, one obtains 
\begin{align}
| \bar{\Phi} \rangle & \equiv \lim_{\underset{\langle A  \rangle  = \text{A}}{\Delta_k \to 0 \,\, \forall k}}  | \Phi \rangle \label{HFBstatelimit} \\
&=  \frac{1}{2}(1 + a^{\dagger}_{v_1} a^{\dagger}_{v_{\bar{1}}}) (1 +  a^{\dagger}_{v_2} a^{\dagger}_{v_{\bar{2}}}) \prod_{h=1}^{(A-2)/2} a^{\dagger}_h a^{\dagger}_{\bar{h}}| 0 \rangle  \, , \nonumber
\end{align}
which is a linear combination of four Slater determinants, one of which has $\text{A}-2=24$ particles (0 particles in the valence shell), two of which have $\text{A}=26$ particles (2 particles in the valence shell) and one that has $\text{A}+2=28$ particles (4 particles in the valence shell). The fact that the limit state $| \bar{ \Phi} \rangle$ can be written as the sum of a {\it finite} number (different from 1) of Slater determinants is remarkable given that $| \Phi \rangle$ can only be expanded over an {\it infinite} sum of Slater determinants as soon as one moves away from the zero-pairing limit.

Given the form of $| \bar{ \Phi} \rangle$ in Eq.~\eqref{HFBstatelimit}, it can easily be checked that the constraint defined through Eq.~\eqref{partnumbconstr} is indeed satisfied
\begin{align}
\langle \bar{ \Phi} | A | \bar{\Phi} \rangle  &= \frac{1}{4}\left[(\text{A}-2) + 2\text{A} + (\text{A}+2) \right] \nonumber \\
&= \text{A} \, ,
\end{align}
even though $| \bar{ \Phi} \rangle$ is {\it not} eigenstate of $A$ in spite of being obtained through the zero-pairing limit. Accordingly, the particle-number variance of $| \bar{ \Phi} \rangle$ is given by
\begin{align}
\text{VAR}_{| \bar{ \Phi} \rangle} &= \frac{1}{4}\left[(\text{A}-2)^2 + 2\text{A}^2 + (\text{A}+2)^2\right] -\text{A}^2 \nonumber \\
&= 2 \, ,
\end{align}
and is thus different from zero.

It is worth noting that $| \bar{ \Phi} \rangle$ does {\it not} correspond to the so-called equal-filling approximation (EFA) that is rigorously formulated on the basis of a mixed-state density matrix operator~\cite{PerezMartin:2008yv} in the sense of statistical quantum mechanics. The limit state $| \bar{ \Phi} \rangle$ obtained in Eq.~\eqref{HFBstatelimit} is a {\it pure} state obtained through the straight resolution of the HFB eigenvalue problem. Its normal density matrix is the same as in the EFA, i.e., $v^2_v = 1/2$ but its anomalous density matrix is also non-zero in the valence shell, i.e., $\kappa_{v\bar{v}}=u_v v_v = 1/2$. As a result, the pairing (i.e., Bogoliubov) contribution to the total energy (Eq.~\eqref{HFBenergy})
\begin{align}
\bar{E}^{\text{B}}_{| \bar{\Phi} \rangle}  & =  
\frac{1}{4}(\overline{v}_{v_{1}v_{\bar{1}}v_{1}v_{\bar{1}}} 
+\overline{v}_{v_{1}v_{\bar{1}}v_{2}v_{\bar{2}}}   \nonumber \\
&\hspace{0.7cm} +\overline{v}_{v_{2}v_{\bar{2}}v_{1}v_{\bar{1}}}  
+\overline{v}_{v_{2}v_{\bar{2}}v_{2}v_{\bar{2}}}) \label{HFBenergylimit2}
\end{align}
is {\it not} zero in the limit state. While  $| \bar{ \Phi} \rangle$  is indeed obtained through a zero-pairing procedure in the sense that the pairing field is strictly driven to zero in the HFB eigenvalue problem (Eq.~\eqref{eq:hfb_equation}), the pairing energy of the limit state is not zero due to the remaining non-zero anomalous density matrix within the valence shell.

Eventually, the above analysis provides several key insights. When driving the spherical HFB state associated with $^{26}$O towards its zero-pairing limit $| \bar{ \Phi} \rangle$, one observes that
\begin{itemize}
\item the limit state $| \bar{ \Phi} \rangle$ carrying A particles on average is mathematically well-defined and takes the form of a linear combination of a {\it finite number}, i.e., 4, of Slater determinants that do not all carry the physical number of particles. As a result, the limit state is not an eigenstate of the particle-number operator.
\item There exists a non-zero lower bound\footnote{The fact that it is indeed a lower bound is proven in App.~\ref{APPvariance}.} to the particle-number variance that can actually be achieved within the manifold of spherical HFB states constrained to carry 26 nucleons (18 neutrons), i.e., 
\begin{equation}
\text{VAR}_{| \Phi \rangle} \geq \text{VAR}_{| \bar{ \Phi} \rangle}=2 \ .
\end{equation}
\item Consistently with this non-zero particle-number variance, the limit state carries a non-zero pairing energy even though it is obtained by diagonalizing an HFB matrix in which the pairing field is vanishing. 
\item As a consequence of Eq.~\eqref{limit4}, the off-diagonal norm overlap associated with $| \bar{ \Phi} \rangle$ (Eq.~\eqref{normoverlap}) is 0 at $\varphi=\pi/2$. While the particle-number projection (PNP) on the value $\text{A}$ is well-defined given that a component with the targeted number of particles does enter $| \bar{ \Phi} \rangle$ according to Eq.~\eqref{HFBstatelimit}, it requires numerical care given that for $d_v=4$ the computation of two-body observables requires a fine-tuned treatment of a zero-over-zero.
\item By virtue of Eqs.~\eqref{limit} and~\eqref{limit3}, the lowest quasi-particle energy  solution of Eq.~\eqref{eq:hfb_equation} fulfills\footnote{The analytical proof of Eq.~\eqref{limit6} relies on using the BCS expression $E_k=\sqrt{(\epsilon_k - \lambda)^2 + \Delta_k^2}$ that is known to be a good approximation of HFB quasi-particle energies, except for s states near the threshold~\cite{doba96a}.}
\begin{align}
\lim_{\underset{\langle A  \rangle  = \text{A}}{\Delta_k \to 0 \,\, \forall k}} \text{Min}_\mu E_{\mu} &=  0 \, , \label{limit6}
\end{align}
such that Eq.~\eqref{fermigap} does not apply anymore in the zero-pairing limit. Thus, the BMBPT expansion based on Option 1 becomes ill-defined in the limit given that zero energy denominators enter; even if the reference energy $\bar{E}_{| \Phi(0) \rangle}$ is well-defined and contains a non-zero Bogoliubov contribution, the second-order correction $E^{(2)}_{| \Phi(0) ; \{E_{\mu}(0)\}\rangle}$ (Eq.~\eqref{BMBPTcorrection2}) diverges. Obviously, no such problem arises in closed-shell nuclei ($o_v=1$) given that BMBPT safely reduces to standard HF-MBPT~\cite{Tichai:2018mll,Arthuis:2018yoo} in this case, with the lowest two quasi-particle excitation converging towards the non-zero particle-hole gap at the Fermi energy.
\end{itemize}

\subsection{General case}
\label{general}

Let us now consider the general open-shell case characterized by $0<o_v<1$ and $o_v \neq 1/2$\footnote{As will be shown below, the case $o_v = 1/2$ must be treated separately such that it was not only a question of convenience to cover it first in Sec.~\ref{specific}.}. The valence shell gathers $p_v=d_v/2$ pairs of conjugated states generically denoted as $(v,\bar{v})$ and presently specified as $(v_1,v_{\bar{1}}), \ldots, (v_{p_v},v_{{\bar{p}_v}})$.  

The zero-pairing limit of hole and particle states works as before such that $\text{A}-a_v$ particles eventually occupy hole states. As for the valence shell, one needs to fit $a_v$ particles in $d_v$ degenerate states characterized by identical occupations $v^{2}_{v_k}\equiv v^{2}_{v}$. To do so, the identity
\begin{align}
a_v &=  2\sum_{k=1}^{p_v} v_{v_k}^2 = d_v \, v^{2}_{v} \, , \label{occuplimit}
\end{align}
must be fulfilled in the zero-pairing limit in order to ensure that
\begin{align}
v^{2}_{v} &= o_v \, .  \label{occuplimit2}
\end{align}
The only way to satisfy Eq.~\eqref{occuplimit2} requires now that $\epsilon_{v} - \lambda$ and $\Delta_{v}$ go to 0 in a strictly proportional fashion, i.e., that
\begin{equation}
\lim_{\underset{\langle A  \rangle  = \text{A}}{\Delta_{v} \to 0}} \left|\frac{\Delta_{v}}{\epsilon_{v} - \lambda}\right| = \gamma \, , \label{limit7}
\end{equation}
with $\gamma$ a non-zero real number. In fact, this property is indeed consistent with Eq.~\eqref{occuplimit2} under the condition that
\begin{equation}
\gamma  = \frac{2\sqrt{o_v(1-o_v)}}{|1- 2 o_v|} \, . \label{propconstant}
\end{equation}
One observes that $\gamma$ is ill-defined for $o_v=1/2$, which reflects the fact that Eq.~\eqref{limit7} is inappropriate in that case and must be replaced by Eq.~\eqref{limit3}, i.e., when $o_v=1/2$ one must rather make the hypothesis that 
\begin{equation}
\lim_{\underset{\langle A  \rangle  = \text{A}}{\Delta_{v} \to 0}} \left|\frac{\Delta^n_{v}}{\epsilon_{v} - \lambda}\right| = \gamma' \, , \label{limit8}
\end{equation}
for some real number $n>1$.

With this at hand, one eventually obtains
\begin{align}
| \bar{\Phi} \rangle & \equiv \lim_{\underset{\langle A  \rangle  = \text{A}}{\Delta_k \to 0 \,\, \forall k}} | \Phi \rangle \label{HFBstatelimit2} \\
&=  \prod_{k=1}^{p_v} (\sqrt{1-o_v} + \sqrt{o_v} \, a^{\dagger}_{v_{k}} a^{\dagger}_{v_{\bar{k}}}) \prod_{h=1}^{(A-a_v)/2} a^{\dagger}_h a^{\dagger}_{\bar{h}}| 0 \rangle \, . \nonumber
\end{align}
Thus, the HFB state carrying A particles on average is well-defined in the zero-pairing limit and takes the form of a linear combination of a {\it finite} number, i.e., $2^{p_v}$, of Slater determinants. Again, the fact that $| \bar{ \Phi} \rangle$ is a finite sum of Slater determinants is remarkable. Among the $2^{p_v}$ Slater determinants, $\binom{b}{p_v}$ of them carry $B(b)=\text{A}-a_v+2b$ particles\footnote{The number of particles carried by the Slater determinants thus ranges from $\text{A}-a_v$ to $\text{A}+ (d_v-a_v)$. The total number of summed Slater determinants is indeed $\sum_{b=0}^{p_v} \binom{b}{p_v} = 2^{p_v}$.}, with the integer $b$ ranging from $0$ to $p_v$. It is easy to see from Eq.~\eqref{HFBstatelimit2} that the weight of each Slater determinant carrying $B(b)$ particles is equal to $o^{b}_v(1-o_v)^{p_v-b}$. 

Given the form of $| \bar{ \Phi} \rangle$, it can first be checked\footnote{Identities~\eqref{binomial1} and~\eqref{binomial2} provided in App.~\ref{formulae} are employed to derive Eq.~\eqref{averGEN} while the additional identity~\eqref{binomial3} is necessary to derive Eq.~\eqref{varGEN}. Similar analytical results could be derived for higher moments of $A$ at the price of considering higher derivatives of Newton's binomial formula.} that the constraint defined through Eq.~\eqref{partnumbconstr} is indeed satisfied in the zero-pairing limit
\begin{align}
\langle \bar{ \Phi} | A | \bar{\Phi} \rangle  &= \sum_{b=0}^{p_v} \binom{b}{p_v} o^{b}_v(1-o_v)^{p_v-b} (\text{A}-a_v+b) \nonumber \\
&= (\text{A}-a_v)\sum_{b=0}^{p_v} \binom{b}{p_v} o^{b}_v(1-o_v)^{p_v-b} \nonumber \\
&\phantom{=} + 2 o_v \sum_{b=1}^{p_v} \binom{b}{p_v} b \, o^{b-1}_v(1-o_v)^{p_v-b}\nonumber \\
&= \text{A}-a_v + 2 o_v  p_v \nonumber \\
&= \text{A} \, , \label{averGEN}
\end{align}
where $o_v=a_v/d_v$ and $p_v=d_v/2$ were eventually used. Similarly, the particle-number variance is obtained after a long but straightforward calculation as
\begin{align}
\text{VAR}_{| \bar{ \Phi} \rangle} &= \sum_{b=0}^{p_v} \binom{b}{p_v} o^{b}_v(1-o_v)^{p_v-b} (\text{A}-a_v+b)^2 \nonumber \\
&\phantom{=} - \text{A}^2 \nonumber \\
&=  2 a_v (1-o_v)\, , \label{varGEN}
\end{align}
and constitutes a lower bound as proven in App.~\ref{APPvariance}. Last but not least, a non-zero pairing contribution
\begin{align}
\bar{E}^{\text{B}}_{| \bar{ \Phi} \rangle}  & =  o_v(1-o_v)\sum_{kl=1}^{p_v}
\overline{v}_{v_{k}v_{\bar{k}}v_{l}v_{\bar{l}}}  \, , \label{HFBenergylimit3}
\end{align}
to the total HFB energy is once again obtained in the limit given that the anomalous density matrix is non-zero within the valence shell and equal, for each canonical pair $(v,\bar{v})$, to $\kappa_{v\bar{v}}= u_v v_v = \sqrt{o_v(1-o_v)}$.

From a general perspective, the present analysis demonstrates that HFB theory does {\it not} reduce to HF even when the pairing field is driven to zero in the HFB Hamiltonian matrix. 

\subsection{Illustrative examples}
\label{examples}

Given the above analysis, typical examples can be discussed on the basis that the expected, i.e., {\it text-book}, sequence of shells is indeed obtained in each case in the zero-pairing limit. This assumption can only be checked \textit{a posteriori} from an actual numerical calculation associated with a code characterized by a certain set of constrained/relaxed symmetries. This important aspect will be scrutinized in Sec.~\ref{Numresults}. Relying for now on the text-book sequence of shells, the following results can reasonably be anticipated.
\begin{itemize}
\item The first typical example, already discussed in Sec.~\ref{specific}, is $^{26}$O. Based on a text-book spherical canonical spectrum, this semi-magic nucleus corresponds to having $a_v=2$ and $d_v=4$, and thus $o_v=1/2$. Inserting these numbers into Eq.~\eqref{varGEN}, the minimal particle-number variance reached in the zero-pairing limit is indeed $\text{VAR}_{| \bar{ \Phi} \rangle} = 2$.
\item A similar but slightly different case relates to the semi-magic $^{44}$Ca nucleus whose naive filling based on a text-book spherical canonical spectrum corresponds to putting 4 particles in the $\text{f}_{7/2}$ shell, i.e., $a_v=4$ and $d_v=8$, thus also leading to $o_v=1/2$. As a result, the minimal particle-number variance obtained in the zero-pairing limit is $\text{VAR}_{| \bar{ \Phi} \rangle} = 4$. 
\item The text-book naive filling of the semi-magic $^{18}$O nucleus corresponds to putting 2 particles in the spherical $\text{d}_{5/2}$ shell, i.e., $a_v=2$ and $d_v=6$, thus leading to $o_v=1/3$. This eventually provides the zero-pairing particle-number variance $\text{VAR}_{| \bar{ \Phi} \rangle} = 8/3$.  
\item One can focus next on the semi-magic $^{22}$O nucleus whose naive filling corresponds to putting 6 particles in the same $\text{d}_{5/2}$ shell, i.e., $a_v=6$ and $d_v=6$, thus leading to $o_v=1$ and a zero minimal variance $\text{VAR}_{| \bar{ \Phi} \rangle} = 0$ as expected for a spherical closed-subshell system. 
\item Considering an even-even doubly open-shell nucleus, e.g., $^{240}$Pu, the unconstrained HFB minimum is typically obtained for a deformed configuration, which is even more so true when the pairing is decreased. Given that the associated canonical spectrum only retains Kramers two-fold degeneracy, the zero-pairing limit state necessarily takes the form of a deformed closed-shell Slater determinant with a zero particle-number variance $\text{VAR}_{| \bar{ \Phi} \rangle} = 0$.
\item Considering the odd-even neighbor, i.e., $^{241}$Pu, the naive filling corresponds to putting 1 particle in a doubly-degenerate valence shell, i.e., $a_v=1$ and $d_v=2$, thus leading to $o_v=1/2$ and to a non-trivial HFB state with $\text{VAR}_{| \bar{ \Phi} \rangle} = 1$. 
\end{itemize}

\begin{figure*}
\scalebox{0.86}{\includegraphics{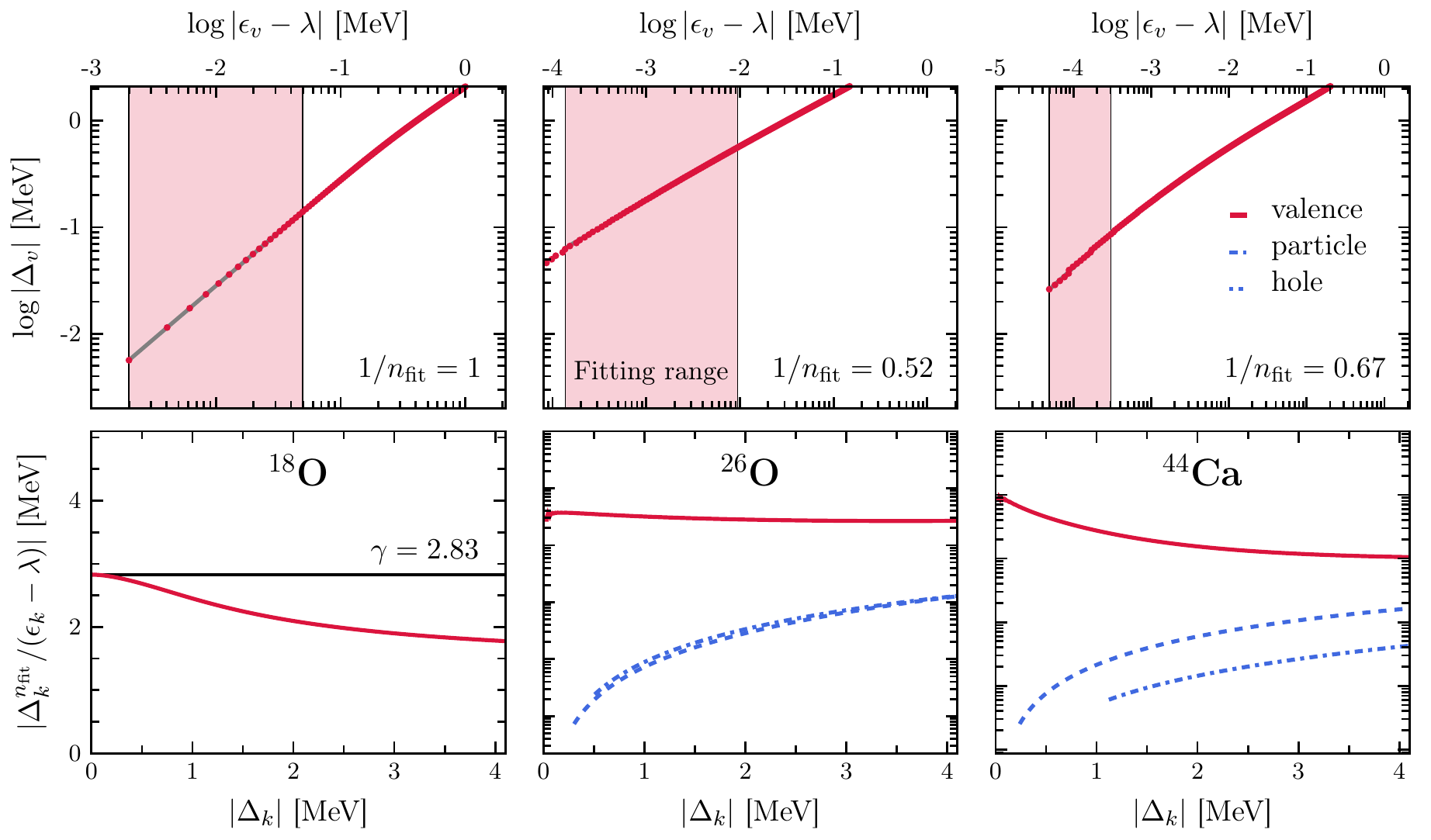}}
\caption{(Color online) Results of constrained HFB calculations of $^{18}$O (left column), $^{26}$O (center column) and $^{44}$Ca (right column). Top row: log-log plot of the valence shell canonical pairing gap $|\Delta_{v}|$ against $|\epsilon_{v} - \lambda|$. The slope $1/n$ of the curve in the limit $\Delta_{v} \rightarrow 0$ (see Eqs.~\eqref{limit7}-\eqref{limit8}) is extracted through a numerical fit. Bottom row: $|\Delta^n_{k}/(\epsilon_{k} - \lambda)|$ as a function of $|\Delta_{k}|$ for the valence shell and for the particle (hole) shell just above (below). The power $n$ employed corresponds to the value extracted in the associated top panel.
\label{fig:limitcharacterization}}
\end{figure*}

\begin{figure*}
\includegraphics[width=1.0\textwidth]{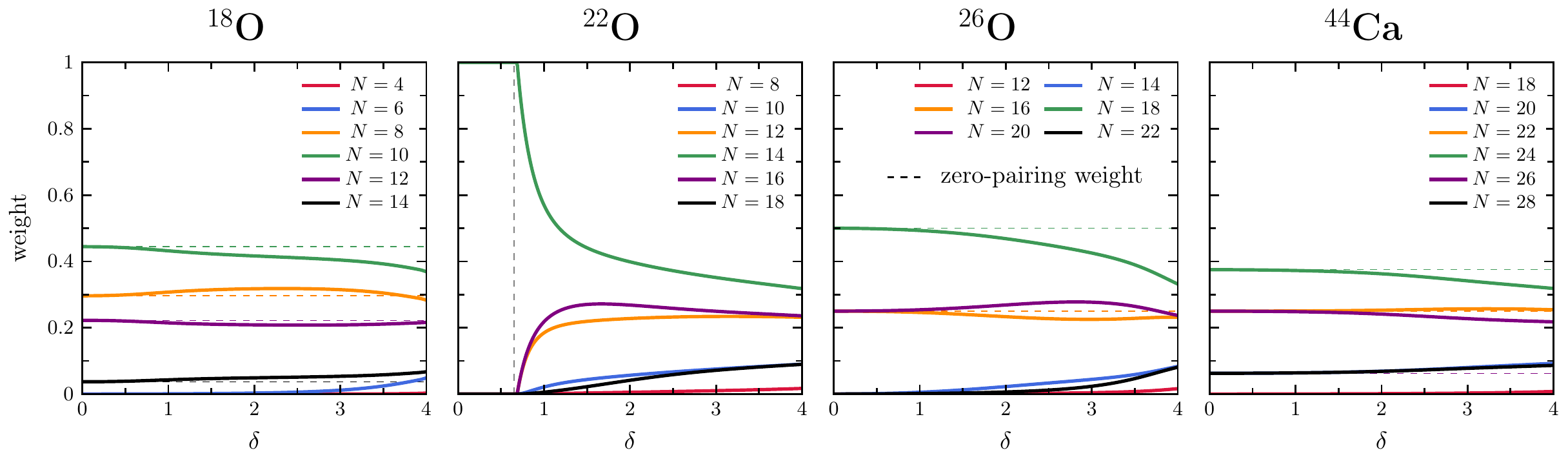}
\caption{Weights of the Slater determinants associated with a given particle number (solid lines) making up the constrained HFB state as a function of $\delta$. The weights are obtained via PNP after variation calculations. The numerical results are compared to the predicted weights in the zero-pairing limit (dashed curves).
\label{weights}}
\end{figure*}

\section{Applications}
\label{Numresults}

In this section, results obtained from constrained HFB and BMBPT calculations based on it, are presented.

\subsection{Numerical set up}
\label{sec:num}

The computations are performed using a realistic nuclear Hamiltonian $H$ derived from chiral effective field theory ($\chi$EFT). The Hamiltonian contains a two-nucleon (2N) interaction derived at next-to-next-to-next-to leading order (N3LO) in the chiral expansion~\cite{Hamil} and evolved down to lower resolution scale ($\alpha = 0.08$ fm$^4$) via a similarity renormalization group (SRG) transformation~\cite{SRGsoft}.

\begin{figure*}
\includegraphics[width=0.8\textwidth]{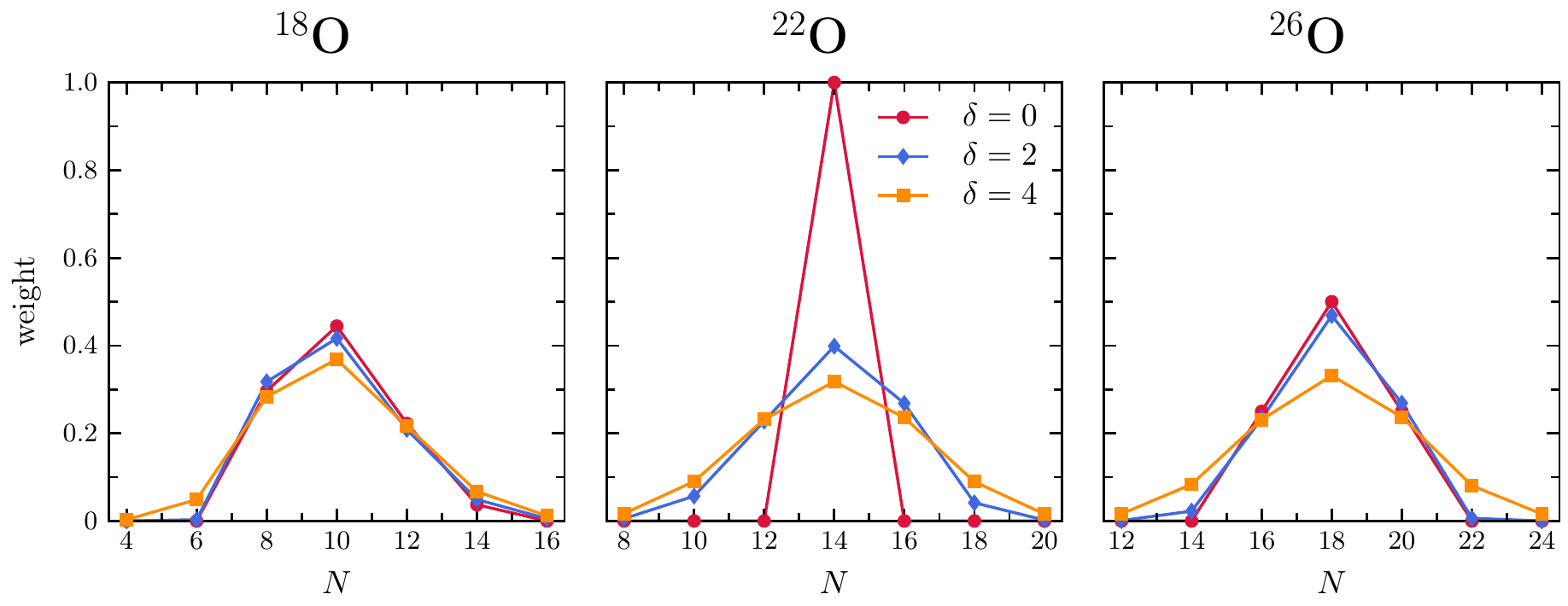}
\caption{Distribution of weights of the good particle-number components of the constrained HFB state in $^{18}$O (left panel) and $^{22}$O (right panel) for $\delta=0,2,4$. \label{weights2}}
\end{figure*}

Two HFB solvers dedicated to \textit{ab initio} calculations, i.e., capable of handling 2N and 3N interactions (either in full or within the normal-ordered two-body approximation \cite{Roth:2011vt}), are presently employed. The first code is restricted to spherical symmetry and is based on the actual diagonalization of the HFB matrix~\cite{Hergert:2009nu}. The second code, named TAURUS$_{\text{vap}}$, solves HFB or variation after particle-number projection (VAPNP) equations for symmetry-unrestricted (real) Bogoliubov quasi-particle states~\cite{bally20a}, thus allowing for spatially deformed solutions. Employing a gradient method, the code can actually solve the variational equations under a large variety of constraints and was recently used to perform first practical calculations~\cite{Bally:2019miu}. 

\begin{figure*}
\includegraphics[width=1.0\textwidth]{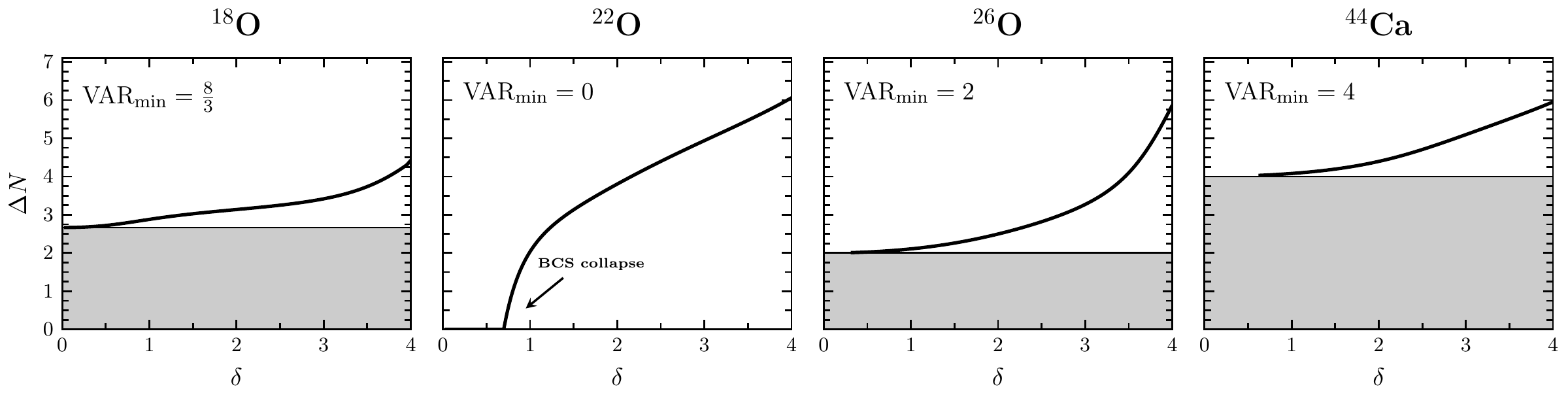}
\caption{Neutron-number variance of the constrained HFB solution for $^{18}$O (left column), $^{22}$O (center-left column), $^{26}$O (center-right column) and $^{44}$Ca (right column) as a function of the constraining parameter $\delta$. In each case, the grey zone materializes the interval of neutron-number variance values that cannot be reached within the manifold of appropriate HFB solutions. The upper limit of the grey zone denotes the predicted value in the zero-pairing limit (Eq.~\eqref{varGEN}) that is provided on each panel as $\text{VAR}_{\text{min}}$. \label{fig:variancevsdelta} }
\end{figure*}

In both codes, one-, two- and three-body operators are represented in the eigenbasis of the spherical harmonic oscillator (SHO) Hamiltonian. In the present calculations, the one-body basis is characterized by a SHO frequency $\hbar \omega = 20$ MeV and includes single-particle states up to $e_{\text{max}} \equiv (2 n + l)_{\text{max}}  = 4$ whereas the two-body basis is built from its tensor product. While realistic \emph{ab initio} calculations typically require to use $e_{\text{max}}=12$ or 14 in mid-mass nuclei to reach convergence with respect to the basis set, calculations performed in a reduced model space are sufficient to investigate the points of present interest.

\subsection{Constrained Hartree-Fock-Bogoliubov}
\label{resultsHFB}

\subsubsection{Characterization of the limit}

For the zero-pairing limit to be analytically meaningful, canonical matrix elements of the pairing field have been predicted in Sec.~\ref{zeropairingSec} to be driven to zero in a specific way when the constraining parameter $\delta$ goes itself to zero. In this context, the half-filled valence-shell case ($o_v=1/2$) had to be explicitly distinguished, i.e., see Eq.~\eqref{limit8} versus Eq.~\eqref{limit7}. In Fig.~\ref{fig:limitcharacterization}, these predictions are tested via numerical calculations of three representative semi-magic nuclei, i.e., $^{18,26}$O and $^{44}$Ca. Under the assumption that they remain spherical all the way down to the zero-pairing limit, the expected text-book shell structures stipulate that these systems all qualify as open-shell nuclei as discussed in Sec.~\ref{examples}.

The top panels of Fig.~\ref{fig:limitcharacterization} display the valence-shell canonical pairing gap $|\Delta_{v}|$ against $|\epsilon_{v} - \lambda|$ in log-log scale. The slope $1/n$ of the curve in the limit $\Delta_{v} \rightarrow 0$ is extracted through a numerical fit. In the general case, i.e., $o_v\neq 1/2$, Eq.~\eqref{limit7} stipulates that both quantities must go to zero in a strictly proportional fashion, i.e., $n=1$. It is indeed what is obtained for $^{18}$O ($o_v=1/3$), thus validating the theoretical prediction. Moving to $^{26}$O and $^{44}$Ca characterized by a half-filled valence shell, the extracted slope is such that $n>1$, also corroborating the prediction. 

Based on the numerical extraction of the parameter $n$ from the top panels, the bottom panels of Fig.~\ref{fig:limitcharacterization} display the ratio $|\Delta^n_{k}/(\epsilon_{k} - \lambda)|$ as a function of $|\Delta_{k}|$ for the valence shell and for the particle (hole) shell just above (below) it. As predicted theoretically, this ratio behaves characteristically in the zero-pairing limit, i.e., it goes to zero for all shells except for the valence shell of open-shell nuclei where it goes to a non-zero value. This behavior is indeed numerically obtained in the three cases. More over, the non-zero limit $\gamma$ was predicted analytically for $o_v\neq 1/2$  (Eq.~\eqref{propconstant}) and is indeed accurately obtained numerically for $^{18}$O. 

\begin{figure*}
\centering
\includegraphics[width=1.0\textwidth]{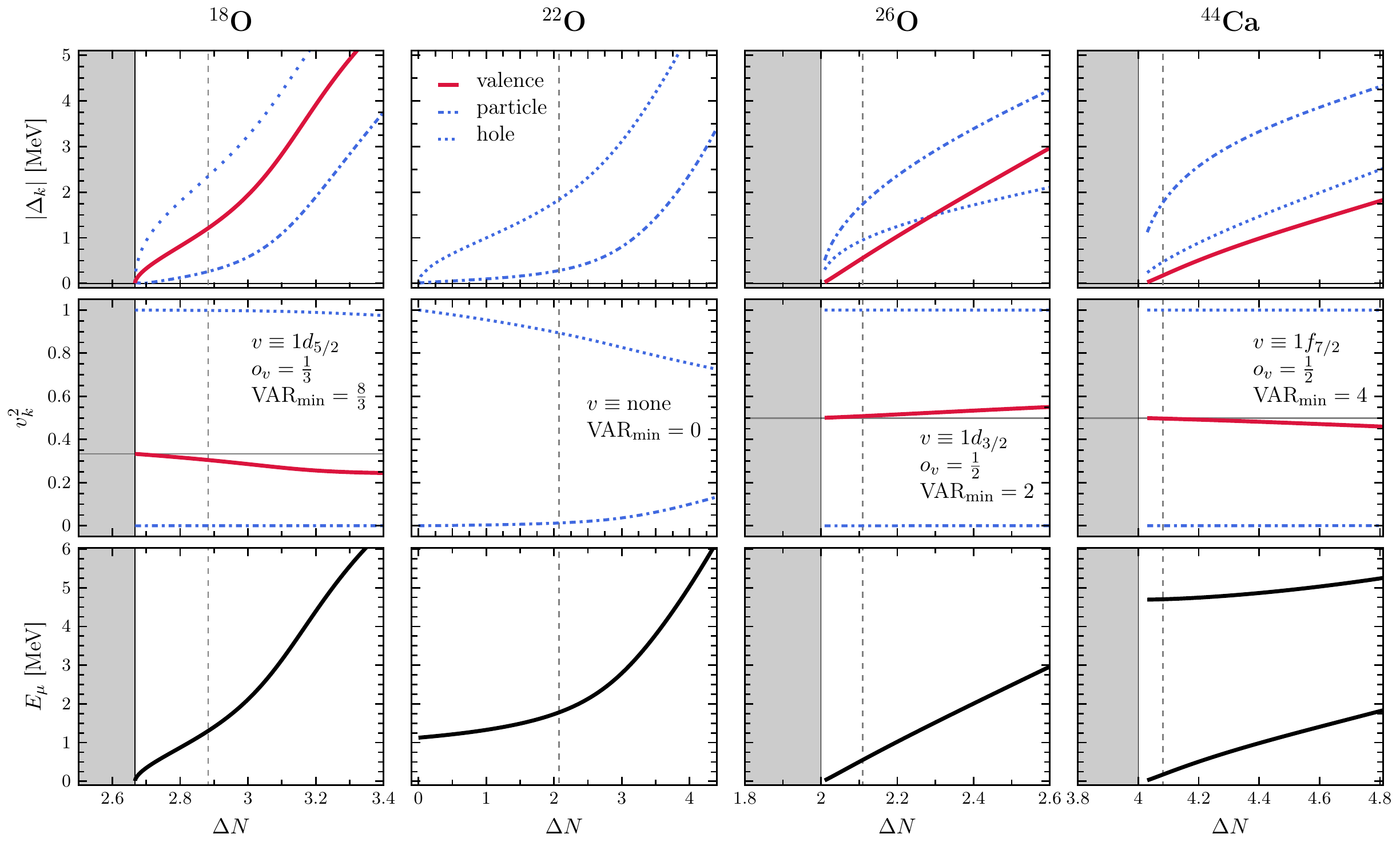}
\caption{(Color online) Results of constrained HFB calculations of $^{18}$O (left column), $^{22}$O (center-left column), $^{26}$O (center-right column) and $^{44}$Ca (right column)  as a function of the neutron-number dispersion. Top row: pairing gaps of neutron canonical states around the Fermi energy. Middle row: average occupation of neutron canonical states around the Fermi energy. Full (red) lines relate to the valence shell whereas the (blue) dotted/dashed-dotted lines relate to the highest hole/lowest particle shells. Bottom row: lowest neutron eigenvalues (i.e., quasi-particle energies) of the HFB matrix. The right-hand limit of the grey zone stipulates the theoretical lower bound of the neutron-particle variance accessible within the manifold of appropriate Bogoliubov states that is reached in the zero-pairing limit (Eq.~\eqref{varGEN}). Horizontal full lines in the center row denote the theoretical value of the valence shell average occupation $o_v$ reached in the zero-pairing limit. Vertical dashed lines characterize the neutron-number dispersion of the unconstrained calculation.
\label{fig:limitdetails}}
\end{figure*}

\subsubsection{Characterization of the limit state}

Now that the analytical premises of the zero-pairing limit have been validated numerically, its consequences on the structure of the HFB solution can be investigated. Based on Eq.~\eqref{HFBstatelimit2}, the limit state $| \bar{\Phi} \rangle$ is predicted to display a specific structure, i.e., to be a linear combination of $2^{p_v}$ Slater determinants. Among them, $\binom{b}{p_v}$ carry $B(b)=\text{A}-a_v+2b$ particles, with $b$ ranging from $0$ to $p_v$, each entering the sum with the weight $o^{b}_v(1-o_v)^{p_v-b}$. This prediction is put to the test in Fig.~\ref{weights} where the weights of each component obtained via particle-number projection after variation (PNPAV) calculations are displayed as a function of $\delta$ for $^{18,22,26}$O and $^{44}$Ca. The numerical weights are compared to the zero-pairing limit prediction, i.e.,  $\binom{b}{p_v} \, o^{b}_v(1-o_v)^{p_v-b}$. 

As visible from the four panels, the HFB state is a linear combination of an infinite number\footnote{In practice, this infinity is of course made finite by the truncation of the one-body Hilbert space to a finite dimension $n_{\text{dim}}$. In this condition, the HFB state mixes Slater determinants spanning the full range of possible (even) particle numbers, i.e., from $0$ to $n_{\text{dim}}$.} of Slater determinants as long as $\delta \neq 0$, albeit with weights quickly decreasing for components moving away from A. In the zero-pairing limit, a qualitatively different structure is obtained, i.e., the linear combination does  collapse to $2^{p_v}$ states carrying the predicted neutron numbers, i.e., from 8 to 14 in $^{18}$O, only 14 in $^{22}$O, from 16 to 20 in $^{26}$O and from 20 to 28 in $^{44}$Ca. Furthermore, the predicted weights are indeed exactly recovered in the limit. 

It must be noted that the single Slater determinant in $^{22}$O is actually reached for a non-zero value of $\delta$. This feature relates to the well-celebrated {\it BCS collapse} and reflects the point at which the pairing strength is too weak to sustain a non-zero pairing field against the finite single-particle gap at the Fermi energy. As visible from Fig.~\ref{weights}, no such pairing collapse occurs in open-shell nuclei. Eventually, the numerical results fully validate the non-trivial structure of the HFB state predicted to be obtained in the zero-pairing limit.

To complement the vision given in Fig.~\ref{weights}, the weights obtained from the PNPAV calculation are displayed differently in Fig.~\ref{weights2} for $^{18,22,26}$O, i.e., the distribution of weights is shown as a function of the particle number for $\delta = 0, 2, 4$. For strong enough pairing, i.e., $\delta = 4$, one recovers the text-book distribution following quite closely a Gaussian distribution in all three cases~\cite{RiSc80}. For a moderate pairing regime, i.e., $\delta = 2$, the distribution may be distorted\footnote{Notice that unconstrained calculations ($\delta=1$) based on the presently used 2N interaction and the omission of any 3N interaction provides too little pairing compared to empirical data.}, e.g., in $^{18}$O. Eventually, the distribution obtained in the zero-pairing limit takes the usual/unusual form for closed-(sub)shell/open-shell isotopes predicted in Sec.~\ref{zeropairingSec}. While the HFB state contains a single non-zero weight associated with the Slater determinant limit in $^{22}$O, the distribution extends over a {\it finite} number of isotopes for open-shell $^{18,26}$O. The distribution over this finite interval is symmetric (asymmetric) in $^{26}$O ($^{18}$O) as a testimony of the naive valence-shell occupation $o_v=1/2$ ($o_v=1/3$). 

\begin{figure}
\includegraphics{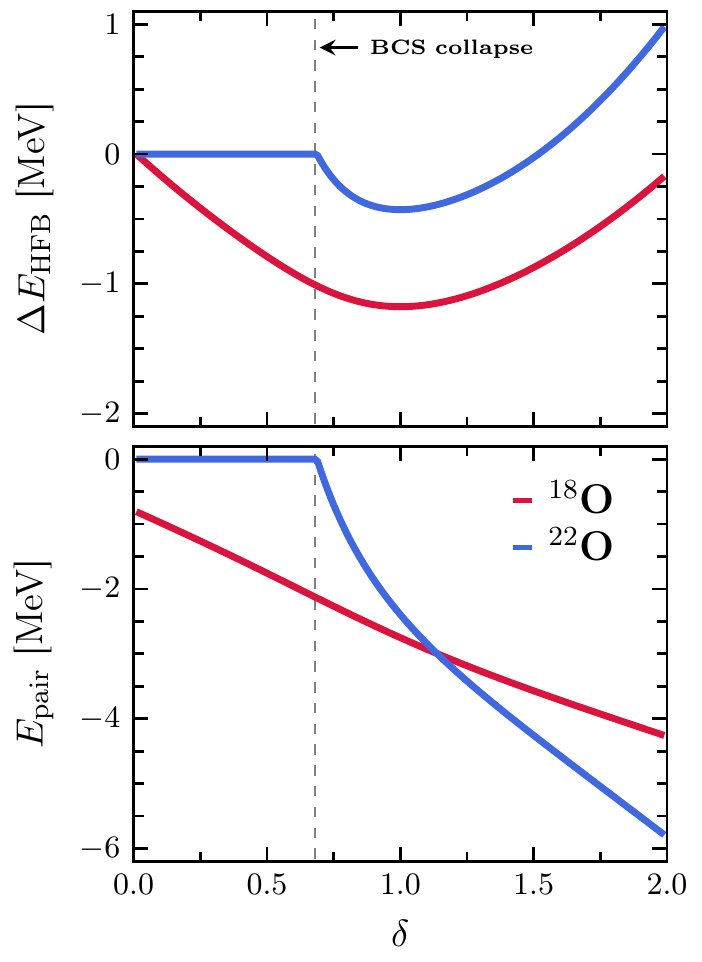}
\caption{(Color online) Results of constrained HFB calculations of $^{18}$O (blue curves) and $^{22}$O (red curves) as a function of the constraining parameter $\delta$ in the HFB matrix. Top panel: total binding energy rescaled to the zero-pairing limit. Bottom panel: contribution of the Bogoliubov term, i.e., pairing energy, to the total binding energy. \label{fig:collapseVSnocollapse} }
\end{figure}

\subsubsection{Particle-number variance}

With the aim to further characterize the zero-pairing limit HFB state, the neutron-number variance is displayed in Fig.~\ref{fig:variancevsdelta} as a function of the constraining parameter $\delta$ for $^{18,22,26}$O and $^{44}$Ca. As expected, the closed-subshell Slater determinant describing $^{22}$O in this limit exhibits a zero neutron-number variance. Contrarily, a non-trivial HFB state displaying a non-zero neutron-number variance $\text{VAR}_{\text{min}}\equiv \text{VAR}_{| \bar{\Phi} \rangle}$ characterizes the three open-shell nuclei. As anticipated, $\text{VAR}_{\text{min}}$ acts as a minimum along the constraining path whose numerical value corresponds in all cases to the one predicted through Eq.~\eqref{varGEN}. 

One must once again note that the zero particle-number variance is reached for a non-zero value for $\delta$ in $^{22}$O, whereas no such pairing collapse occurs in open-shell nuclei.

\begin{figure}
\includegraphics{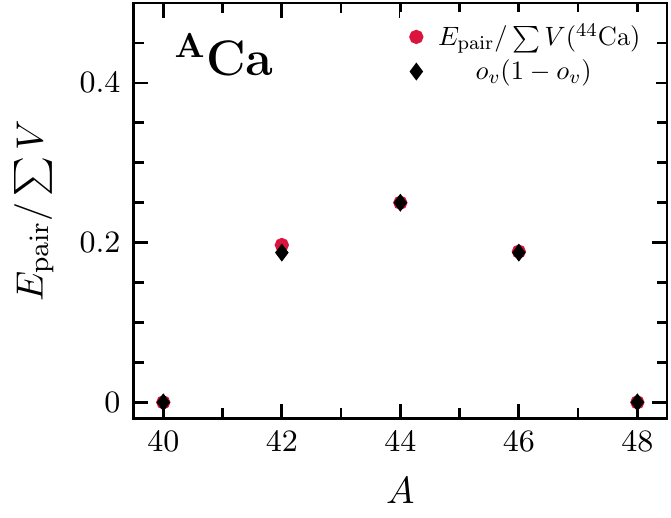}
\caption{(Color online) Rescaled pairing energy (see text) in the zero-pairing limit for $^{40-48}$Ca compared to the analytical prediction $o_v(1-o_v)$. To compute the latter, $o_v$ is taken to be the value obtained for each nucleus on the basis that the neutron f$_{7/2}$ shell indeed acts as the valence shell in the zero-pairing limit. \label{fig:pairingenergy} }
\end{figure}

\subsubsection{Spectroscopic quantities}

The HFB state $| \bar{\Phi} \rangle$ reached in the zero-pairing limit is further scrutinized in Fig.~\ref{fig:limitdetails} where several key quantities are displayed as a function of the neutron-number variance for $^{18,22,26}$O and $^{44}$Ca. Whereas canonical pairing gaps near the Fermi energy are visible in the top panels, associated canonical single-particle occupations are shown in the middle panels. While pairing gaps are driven to zero when the neutron-number variance reaches  $\text{VAR}_{\text{min}}$, single-particle occupations converge to the expected values for all four nuclei, e.g., the valence-shell occupation smoothly attains the naive-filling value associated with a text-book spherical canonical spectrum, e.g., $v^2_v = o_v = 1/3$ in $^{18}$O.

In the bottom panel, quasi-particle energies, i.e., eigenvalues of the constrained HFB equation are displayed below $6$\,MeV. The lowest quasi-particle energy reaches a non-zero value for the closed-subshell nucleus $^{22}$O in the zero-pairing limit. Contrarily, the lowest quasi-particle energy does go to zero in $^{18,26}$O and $^{44}$Ca. While the former characteristic reflects the presence of the finite particle-hole gap at the Fermi energy in $^{22}$O, the latter is a fingerprint of the open character of the valence shell in $^{18,26}$O and  $^{44}$Ca. The two different behaviors have decisive consequences for the application of BMBPT in the zero-pairing limit as discussed in Sec.~\ref{resultsBMBPT} below.

\begin{figure}
\scalebox{0.72}{\includegraphics{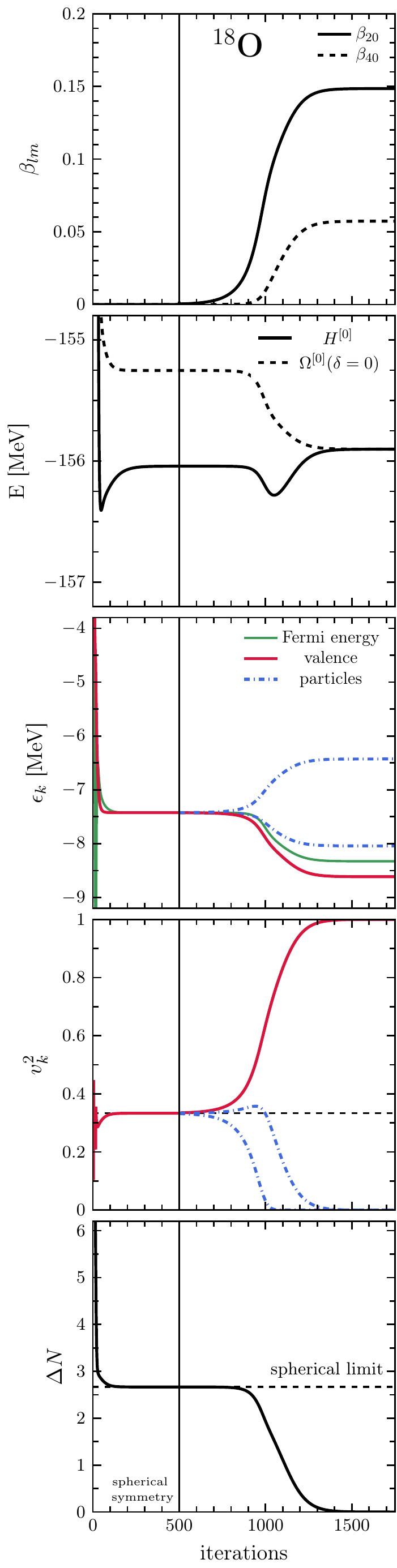}}
\caption{(Color online) Results of constrained HFB calculations of $^{18}$O in the zero-pairing limit ($\delta =0$) as a function of the number of iterations. Top panel: axial quadrupole ($\beta_{20}$) and hexadecapole ($\beta_{40}$) deformations. Second-to-top panel: total constrained grand potential and Hamiltonian expectation values. Middle panel: canonical single-particle spectrum. Second-but-last panel: canonical single-particle occupations. Bottom panel: particle-number variance. Horizontal dashed lines stipulate the theoretical limits for a spherical solution with a text-book single-particle spectrum, i.e., a neutron d$_{5/2}$ valence shell. The vertical full line denotes the point at which the converged spherical solution is constrained to deformation $\beta_{20} = 0.001$ during one iteration.
\label{fig:symmetrychange}}
\end{figure}

\subsubsection{Binding energy}

To complement the numerical analysis, Fig.~\ref{fig:collapseVSnocollapse} provides the total HFB energy, as well as its pairing (i.e., Bogoliubov) contribution, in $^{18,22}$O as a function of the constraining parameter $\delta$. The behavior is again qualitatively distinct for closed-(sub)shell and open-shell nuclei. In $^{22}$O, the pairing energy goes to zero for a non-zero values of $\delta$, a point at which the total energy is non-differentiable, thus, signaling the sharp transition associated with the BCS collapse\footnote{The sharp transition to the non-superfluid phase is non-physical in finite systems such as atomic nuclei and is known to be an artefact of the HFB/BCS theory. This deficiency is resolved at the VAPNP level.} to the non-superfluid phase. Contrarily, the constrained HFB energy evolves smoothly all the way down to $\delta =0$ in $^{18}$O, point at which its pairing component is still different from zero. 

According to Eq.~\eqref{HFBenergylimit3}, the pairing contribution to the energy is predicted to evolve characteristically when filling a given valence shell in the zero-pairing limit. Under the assumption that the sum of valence-shell interaction matrix elements making up the second factor at play in Eq.~\eqref{HFBenergylimit3} is constant while filling the shell, the pairing energy should be strictly proportional to $o_v(1-o_v)$. This prediction is put to the test in Fig.~\ref{fig:pairingenergy} from $^{40}$Ca till $^{48}$Ca, i.e., when filling up the neutron f$_{7/2}$ shell. Dividing $\bar{E}^{\text{B}}_{| \bar{ \Phi} \rangle}$ by the sum of matrix elements evaluated in $^{44}$Ca, the rescaled pairing energy is compared to $o_v(1-o_v)$, where the latter is evaluated on the basis that the neutron f$_{7/2}$ shell indeed acts as the valence shell in the zero-pairing limit. In spite of the fact that the canonical basis, and thus the sum of matrix elements at play in Eq.~\eqref{HFBenergylimit3}, differs in each nucleus in principle, the rescaled pairing energy closely follows $o_v(1-o_v)$, thus confirming the theoretical prediction.

\begin{figure*}
\centering
\includegraphics[width=1.0\textwidth]{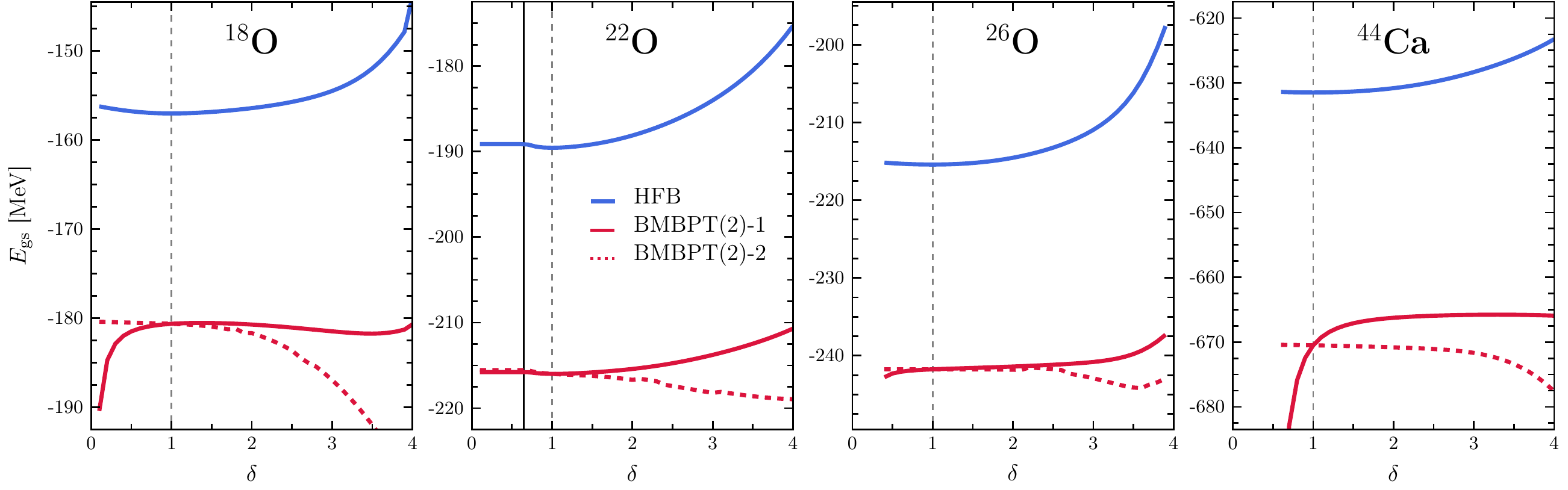}
\caption{(Color online) Potential energy surface of $^{18}$O (left column), $^{22}$O (center-left column), $^{26}$O (center-right column) and $^{44}$Ca (right column)  as a function of the constraining parameter $\delta$. Results are showed for HFB and for the two variants of BMBPT(2) (see text). Vertical dashed lines localize the unconstrained HFB calculation. \label{fig:HFB_BMBPT2} }
\end{figure*}

\subsubsection{Impact of spatial symmetries}

So far, the zero-pairing limit has been investigated in semi-magic nuclei under the hypothesis that the associated solution of the constrained HFB problem remains spherical all the way down to the pairing limit. In the present section, the possibility that the system breaks spherical symmetry, i.e., lowers the constrained Routhian by relaxing spherical symmetry, is investigated. Because it is clear that all doubly open-shell systems do deform in the zero-pairing limit, the situation is more subtle, and thus more informative, in semi-magic nuclei. As a result, $^{18}$O is first used as a primer example.

Figure~\ref{fig:symmetrychange} displays the results of a HFB calculation of $^{18}$O in the zero-pairing limit, i.e., constrained to $\delta =0$, as a function of the number of iterations in the minimization process. Inputting a spherical ansatz at iteration 0, the system evolves freely until iteration 499, at which point it is subjected to an infinitesimal constraint on the axial quadrupole deformation ($\beta_{20} = 0.001$)\footnote{The multipole moments are computed as $\beta_{lm} \equiv 4\pi/[3A(1.2 A^{1/3})^l] \langle \Phi | Q_{lm} | \Phi \rangle $, where $Q_{lm} \equiv r^l Y_{lm}(\hat{\boldsymbol{r}})$ denotes the multipole operator~\cite{Ryssens:2015kga}.} during one iteration, before continuing the unconstrained iteration process until convergence.

The top panel of Fig.~\ref{fig:symmetrychange} shows that the solution remains strictly spherical for the first 499 iterations given that rotational invariance is a {\it self-consistent} symmetry, i.e., the input solution at iteration 0 being spherically symmetric, the invariance cannot be spontaneously broken during the minimization process. Because the solution is provided with an infinitesimal quadrupole deformation at iteration 500, the system can take advantage of deformation for the remaining iteration process. Indeed, $^{18}$O constrained to the zero-pairing limit does deform and converges to a state with non-zero axial quadrupole and hexadecapole deformations.  

The second panel of Fig.~\ref{fig:symmetrychange} confirms that it is indeed advantageous for the system to exploit spatial deformation, i.e., while a fully converged spherical solution is obtained by the time iteration 499 is reached, the system does further lower the constrained Routhian once allowed to deform such that a newly converged solution is obtained by the time iteration 1500 is reached\footnote{While the present calculation is performed in the zero-pairing limit, it is not always advantageous for $^{18}$O to deform when $\delta \neq 0 $. With the present Hamiltonian and for parity-conserving axially-deformed calculations (which is the framework to consider after the perturbation at iteration 500), the minimum of the constrained Routhian is a spherical Bogoliubov state for $0.70 \lesssim \delta \leq 1$, a non-trivial deformed Bogoliubov state for $0.30 \lesssim \delta \lesssim 0.70$ and a deformed Slater determinant for $0 \leq \delta \lesssim 0.3$.}. However, while the constrained Routhian is indeed lower for the deformed configuration, the constrained HFB energy $E_{| \Phi(\delta) \rangle}$ is not. This means that the deformed solution corresponds to an {\it excited} configuration higher in energy than the converged spherical configuration reached before iteration 499 in the zero-pairing limit. 

The fact that the spherical configuration has lower energy than the deformed one is a consequence of their distinct nature as can be understood from the last three panels displaying canonical single-particle energies and occupations as well as the neutron-number variance. Before iteration 499, the spherical solution corresponds to the zero-pairing limit discussed at length for $^{18}$O in previous sections, i.e., a non-trivial HFB state corresponding to a partially filled d$_{5/2}$ valence shell and a neutron-number variance equal to $8/3$. Contrarily, the converged solution obtained after nearly 1500 iterations is a deformed Slater determinant with zero neutron-number variance. From iteration 500 till convergence, the spherical degeneracy of canonical single-particle energies is progressively lifted to give rise to the two-fold Kramers degeneracy. In particular, the d$_{5/2}$ valence shell is split into three pairs of doubly degenerate shells among which the lowest pair becomes gradually fully filled whereas the other two become fully empty. Consequently, the constrained solution transitions in the zero-pairing limit from a spherical open-shell HFB state characterized by $o_v=1/3$ to a deformed closed-shell Slater determinant characterized by $o_v=1$. This change of structure indeed has a marked impact on the energetic of the system. While the spherical HFB solution benefits from a non-zero pairing contribution responsible for the lowering of the constrained energy compared to the constrained Routhian, it is not the case for the deformed Slater determinant for which both quantities are equal. The net result is that the constrained HFB energy is eventually lower for the spherical configuration than for the deformed one. 

However, it happens that this behavior is also a consequence of the specificities of the perturbation imposed at iteration 500, in particular of its remaining symmetries. Indeed, while applying a constraint on $\beta_{20}$ for one iteration opens up the possibility for the system to deform later on, it does so only for parity-conserving axial deformations.
When considering a fully symmetry-unrestricted ansatz\footnote{Still preserving the separation between neutrons and protons.}, the HFB reference state obtained at $\delta=0$ is a deformed Slater determinant with a total energy below the one of the spherical solution even if only by a few keV.

Moreover, systematic calculations over the full set of even-even sd-shell nuclei performed in the zero-pairing limit show that, although a non-trivial HFB solution exists for semi-magic open-shell nuclei when enforcing spherical symmetry, the actual symmetry-unrestricted ground-state is systematically provided by a Slater determinant with zero particle-number variance\footnote{For odd-even nuclei, the deformed even-number parity solution remains a non-trivial HFB state in the zero-pairing limit. Running $^{19}$O as an example, its spherically-symmetric solution is a non-trivial HFB state associated with the spherical d$_{5/2}$ valence shell ($a_v=3$, $d_v=6$, $o_v=1/2$) and a neutron-number variance equal to 3 whereas the symmetry-unrestricted solution is a deformed HFB state with neutron-number variance equal to 1. If one where to search for a odd-number parity solution associated with one quasi-particle excitation, $^{19}$O would converge in the zero-pairing limit to a deformed Slater determinant breaking time-reversal symmetry, and thus Kramers degeneracy, carrying zero neutron-number variance.}, either spherical or deformed depending on the nucleus considered. 

%To realize this, systematic calculations over the full set of even-even sd-shell nuclei have been performed in the zero-pairing limit. Except if the system is a spherical closed-(sub)shell, and is thus described by a spherical Slater determinant, the minimum of the constrained Routhian is always obtained for a deformed Slater determinant. Furthermore, except for $^{18}$O, the constrained HFB energy is also systematically lower for the deformed configuration than for the spherical one with the presently used Hamiltonian. Thus, although a non-trivial HFB solution exists in the zero-pairing limit for semi-magic open-shell nuclei when enforcing spherical symmetry, the actual symmetry-unrestricted ground-state is systematically provided by a deformed Slater determinant with zero particle-number variance\footnote{For odd-even nuclei, the deformed even-number parity solution remains a non-trivial HFB state in the zero-pairing limit. Running $^{19}$O as an example, its spherically-symmetric solution is a non-trivial HFB state associated with the spherical d$_{5/2}$ valence shell ($a_v=3$, $d_v=6$, $o_v=1/2$) and a neutron-number variance equal to 3 whereas the symmetry-unrestricted solution is a deformed HFB state with neutron-number variance equal to 1. If one where to search for a odd-number parity solution associated with one quasi-particle excitation, $^{19}$O would converge in the zero-pairing limit to a deformed Slater determinant breaking time-reversal symmetry, and thus Kramers degeneracy, carrying zero neutron-number variance.}.

\subsection{Bogoliubov many-body perturbation theory}
\label{resultsBMBPT}

Based on spherical HFB reference states, BMBPT(2) calculations have been performed as a function of the pairing constraint using the numerical code whose first results were reported in Ref.~\cite{Tichai:2018mll}. The two options regarding the definition of the unperturbed Hamiltonian $\Omega_{0}$ discussed in Sec.~\ref{BMBPTconstrained} have been tested. The associated potential energy surfaces (PES) are displayed in Fig.~\ref{fig:HFB_BMBPT2} for four nuclei along with the first order, i.e., HFB, one. Away from the unconstrained HFB minimum, the average particle number receives a non-zero contribution at second order such that the reference value must be iteratively reajusted in order for the sum of both contributions to match the physical value A. In Ref.~\cite{Demol:2020mzd}, a so-called {\it a posteriori} correction was shown to provide an excellent approximation to this costly readjustment method. This {\it a posteriori} correction is presently utilized.

Focusing on Option 1, i.e., BMBPT(2)-1 results, one observes that the PES essentially retains the memory of the HFB one, albeit the several tens of MeV of added correlation energy. Looking closer, one however remarks that the PES becomes markedly different in $^{18}$O, $^{26}$O and $^{44}$Ca as the zero-pairing limit is approached. Because the lowest quasi-particle energy of the constrained HFB spectrum  $\{E_{\mu}(\delta)\}$ goes to zero in open-shell nuclei as the limit is reached, the second-order correction $E^{(2)}_{| \Phi(\delta) \rangle; \{E_{\mu}(\delta)\}}$ (Eq.~\eqref{BMBPTcorrection2}) diverges as $\delta \rightarrow 0$. Obviously, no such problem occurs in $^{22}$O given that BMBPT safely reduces to HF-MBPT in this case, with the lowest two quasi-particle excitation converging towards the non-zero particle-hole gap at the Fermi energy.

Moving to BMBPT-2, the issue arising in the zero-pairing limit is regularized by construction, i.e.,  defining $\Omega_0$ from the unconstrained HFB spectrum $\{E_{\mu}(1)\}$ for all values of $\delta$, the energy denominators at play in the second-order energy correction can never be singular. As a result, none of the PES diverges as $\delta \rightarrow 0$. However, the PES behave at odd with HFB and BMBPT-1 when increasing $\delta$. This behavior relates to the non-canonical term $\breve{\Omega}^{11}_{| \Phi(\delta) \rangle; \{E_{\mu}(1)\}}$ becoming very large, probably leading to a highly diverging BMBPT expansion. Thus, except for the benefit brought by construction in zero-pairing limit, the (non-standard) partitioning associated with BMBPT-2 is not to be trusted in general and shall probably only remain as an academic exercise performed for the sake of the present study.

\section{Conclusions}
\label{conclusions}

The zero-pairing limit of an even-number parity Bogoliubov state solution of the Hartree-Fock-Bogoliubov equation under the constraint to carry a fixed number of particles A on average has been investigated in details, i.e., both analytically and numerically.  This investigation is both of academic interest and of relevance to calculations involving a constraint on a collective variable that directly or indirectly impacts the amount of pairing correlations in the system.

It was demonstrated that the HFB state reaches a mathematically well-defined limit, independently of the closed- or open-shell character of the system. While the HFB state trivially goes to a Slater determinant carrying A particles in closed-(sub)shell systems, it converges in open-shell systems to a specific linear combination of a finite number of Slater determinants, among which only a subset carries the physical particle number A. Consequently, and in spite of being obtained through a zero-pairing limit, the corresponding state carries a non-zero pairing energy and a non-zero particle-number variance acting as the lower bound accessible within the manifold of appropriate HFB states.  From a general perspective, the present analysis demonstrates that HFB theory does {\it not} reduce to HF theory even when the pairing field is driven to zero in the HFB Hamiltonian matrix. 

All the characteristics of the HFB state predicted analytically in the zero-pairing limit have been validated numerically for a selected set of representative nuclei. Calculations were performed on the basis of a realistic two-nucleon interaction derived within the frame of chiral effective field theory but are actually generically valid.  

Eventually, the consequences of taking the zero-pairing of the HFB state on expansion many-body methods built on top of it, e.g., Bogoliubov many-body perturbation theory, have been further illustrated. While BMBPT smoothly goes to standard, i.e., Slater-determinant-based, many-body perturbation theory for closed-(sub)shell systems, it becomes ill-defined for open-shell systems when taking the zero-pairing limit.

It will be interesting to extend the investigation of the zero-pairing limit to the finite-temperature Hartree-Fock-Bogoliubov formalism in the future.

\section{Acknowledgements}

The authors thank J. Ripoche for having numerically identified the lower bound on the particle-number variance for open-shell systems and for pointing it out to them and W. Ryssens for useful discussions. This project is supported by the European Union’s Horizon 2020 research and innovation programme under the Marie Skłodowska-Curie grant agreement No.~839847, the Max Planck Society and the Deutsche Forschungsgemeinschaft (DFG, German Research Foundation) -- Project-ID 279384907 -- SFB 1245. The authors thank Heiko Hergert for sharing his spherical HFB code and Robert Roth for providing them with the interaction matrix elements.

\begin{appendix}

\section{Useful formulae}
\label{formulae}

Newton's binomial formula along with its first and second derivatives with respect to $x$ provide three useful identities
\begin{align}
(x+y)^n &= \sum_{k=0}^{n} \binom{k}{n} \, x^{k} y^{n-k} \, , \label{binomial1} \\
n(x+y)^{n-1} &= \sum_{k=1}^{n} \binom{k}{n} \, k \, x^{k-1} y^{n-k} \, , \label{binomial2} \\
n(n-1)(x+y)^{n-2} &= \sum_{k=2}^{n} \binom{k}{n} \, k(k-1) x^{k-2} y^{n-k} \, . \label{binomial3}
\end{align}

\section{Minimal particle-number variance}
\label{APPvariance}

Given the second-quantized form of $A$, the average particle-number variance associated with a Bogoliubov state is easily obtained via Wick's theorem under the form
\begin{align}
\text{VAR}_{| \Phi \rangle}
  &= \sum_{\alpha\beta} \left( \kappa_{\alpha\beta}^* \kappa_{\alpha\beta} - \rho_{\beta\alpha} \rho_{\alpha\beta}  \right) + \sum_{\alpha} \rho_{\alpha\alpha} \notag \\*
  &= - \text{Tr}[\kappa\kappa^\ast] - \text{Tr}[\rho^2]  + \text{Tr}[\rho] \, , \label{eq:pnvar_rho_kappa}
\end{align}
which is a positive or null quantity. Resorting to the unitarity of the Bogoliubov transformation~\cite{RiSc80}, the identity
\begin{align}
  \label{eq:rho_kappa_trace_relation}
  -\text{Tr}[\kappa \kappa^\ast] + \text{Tr}[\rho^2] - \text{Tr}[\rho]
  &= 0 \, ,
\end{align}
can be proven and added $(2\alpha-1)$ times, with $\alpha$ an arbitrary real number, to Eq.~\eqref{eq:pnvar_rho_kappa} to generate alternative expression
\begin{align}
\text{VAR}_{| \Phi \rangle}
  &\equiv -2\alpha \text{Tr}[\kappa\kappa^\ast] + 2(1-\alpha) (\text{Tr}[\rho] - \text{Tr}[\rho^2]) \, . \label{eq:pnvar_rho_kappa2}
\end{align}
This procedure allows one to vary the proportion with which the terms depending on normal or anomalous density matrices contribute to the particle number variance. Choosing $\alpha = 0$, an expression depending solely on the normal density matrix is conveniently obtained
\begin{align}
\text{VAR}_{| \Phi \rangle}
  &= 2 (\text{Tr}[\rho] - \text{Tr}[\rho^2])  \nonumber \\
  &= 2 \big( \sum_{k} v^2_k - \sum_{kk'} v^2_k v^2_{k'}\big)\, , \label{eq:pnvar_rho}
\end{align}
where the latter expression results from using the canonical basis. 

Next,  the differential form of the particle-number variance under an infinitesimal variation of the canonical pairing field matrix elements is computed
\begin{align}
\delta \text{VAR}_{| \Phi \rangle}
  &= \sum_{kk'>0} \frac{\delta \text{VAR}_{| \Phi \rangle}}{\delta \Delta_{k\bar{k}'}} \delta \Delta_{k\bar{k}'} \nonumber \\
  &= \sum_{kk'>0} \sum_{l>0} \frac{\delta \text{VAR}_{| \Phi \rangle}}{\delta v^2_{l} } \frac{\delta v^2_{l} }{\delta \Delta_{k\bar{k}'}}  \delta \Delta_{k\bar{k}'} \nonumber \\
  &= \sum_{k>0} \frac{(\epsilon_k - \lambda)^2|\Delta_k|}{[(\epsilon_k - \lambda)^2 + \Delta_k^2]^{2}} \, \delta |\Delta_k|. \label{diff}
\end{align}
where the partial derivatives at play have been obtained using both Eqs.~\eqref{eq:pnvar_rho} and~\eqref{partnumbconstr}
\begin{subequations}
\label{partialderivatives}
\begin{align}
\frac{\delta \text{VAR}_{| \Phi \rangle}}{\delta v^2_{l} }  &=   2(1-2v^2_{l})  \nonumber \\
&= \frac{2(\epsilon_l - \lambda)}{\sqrt{(\epsilon_l - \lambda)^2 + \Delta_l^2}}\, ,  \label{partialderivatives1} \\
\frac{\delta v^2_{l} }{\delta \Delta_{k\bar{k}'}}    &=   \frac{(\epsilon_k - \lambda)|\Delta_k|}{2[(\epsilon_k - \lambda)^2 + \Delta_k^2]^{3/2}} \delta_{kk'} \delta_{kl} \, . \label{partialderivatives2}
\end{align}
\end{subequations}
Consequently, the particle-number variance is an increasing function of each of the canonical pairing gap matrix elements and takes its minimum value when all these pairing  gap matrix elements go to zero, i.e., in the zero-pairing limit. While the actual value of the particle-number variance reached in the zero-pairing limit does depend on the situation, i.e., on the particle number A and the (symmetry of the) canonical spectrum, as discussed at length in the body of the paper, this value acts as a lower bound within the manifold of HFB states characterized by a given symmetry and a given average particle number A.

\end{appendix}

\bibliographystyle{apsrev4-1}
\bibliography{bib_nucl}

\end{document}